\documentclass[journal]{IEEEtran}
\usepackage{graphicx}
\usepackage{xcolor} % for coloring the text
\usepackage{soul} % for highlighting the text
\usepackage{units} % for inserting units with spacing
\usepackage{array} %centering columns in a table with predefined column widths
\usepackage{amsmath} % use non-italic text
\usepackage{enumitem} % to allow left indentation of list items
\usepackage{tabularx} % to control the table width more precisely
\usepackage{listings}
\usepackage{multirow}
\usepackage{float}
\usepackage{wrapfig}
\newcolumntype{L}{>{\raggedright\arraybackslash}X} % multi-line rows
\newcolumntype{P}[1]{>{\centering\arraybackslash}p{#1}}
\newcolumntype{C}[1]{>{\centering\arraybackslash}m{#1}}

\usepackage{amssymb, amsmath} 

\usepackage{listings}
\usepackage{color}

\definecolor{dkgreen}{rgb}{0,0.6,0}
\definecolor{gray}{rgb}{0.5,0.5,0.5}
\definecolor{mauve}{rgb}{0.58,0,0.82}

\begin{document}

\title{ALPINE: Analog In-Memory Acceleration with Tight Processor Integration for Deep Learning}

\author{
    \IEEEauthorblockN{
        Joshua Klein$^1$, Irem Boybat$^2$, Yasir Qureshi$^1$, Martino Dazzi$^2$, Alexandre Levisse$^1$, \\ Giovanni Ansaloni$^1$, Marina Zapater$^3$, Abu Sebastian$^2$, David Atienza$^1$
    } \\
    \IEEEauthorblockA{
        $^1$Embedded Systems Laboratory (ESL), École Polytechnique Fédérale de Lausanne (EPFL), Lausanne, Switzerland
        \\ $^2$IBM Research Europe, Ruschlikon, Switzerland,
        \\ $^3$HEIG-VD, Yverdon-les-Bains, Switzerland
        \\
    }
    % \and
    % \IEEEauthorblockN{
    %     Person 1, Person 2, Person 3
    % }
    % \IEEEauthorblockA{
    %     Over the rainbow
    % }
    % Need: EPFL, HES-SO, IBMZ
}

\maketitle
\begin{abstract}
Analog in-memory computing (AIMC) cores offers significant performance and energy benefits for neural network inference with respect to digital logic (e.g., CPUs). AIMCs accelerate matrix-vector multiplications, which dominate these applications’ run-time. However, AIMC-centric platforms lack the flexibility of general-purpose systems, as they often have hard-coded data flows and can only support a limited set of processing functions. With the goal of bridging this gap in flexibility, we present a novel system architecture that tightly integrates analog in-memory computing accelerators into multi-core CPUs in general-purpose systems. We developed a powerful gem5-based full system-level simulation framework into the gem5-X simulator, ALPINE, which enables an in-depth characterization of the proposed architecture. ALPINE allows the simulation of the entire computer architecture stack from major hardware components to their interactions with the Linux OS. Within ALPINE, we have defined a custom ISA extension and a software library to facilitate the deployment of inference models. We showcase and analyze a variety of mappings of different neural network types, and demonstrate up to 20.5x/20.8x performance/energy gains with respect to a SIMD-enabled ARM CPU implementation for convolutional neural networks, multi-layer perceptrons, and recurrent neural networks.
\end{abstract}
\IEEEpeerreviewmaketitle

\section{Introduction}
%%% INTRODUCTION %%%

Deep neural networks (DNNs) have revolutionized the state-of-the-art in a wide range of AI applications ranging from computer vision to natural language and speech processing. DNNs are composed of multiple consecutive layers, and their ability to address tasks often increases with the number and size of layers. Today,  modern DNNs are composed of hundreds of layers and millions of weights, requiring massive amounts of computation and memory \cite{Y2016heCVPR, Y2017VaswaniNeurIPS}. In the quest for achieving higher accuracy and throughput, the AI domain suffers from constant changes in the type and nature of the DNNs. 

One of the main contributing factors to the expansion of DNNs has been the introduction of more powerful CPUs and GPUs. However, system solutions based on CPU-GPU integration struggle to meet the efficiency requirements of the edge domain, where the extreme data-intensive nature of DNNs mandates efficient storing, accessing and processing of large amounts of data. Efficiency gains can be achieved by lowering data precision (e.g., from 64/32-bit floating point to 8/4-bit integer)~\cite{Y2020ChoquetteHCS}.  Yet, the time and energy for accessing the data still dominate over the data processing~\cite{Y2021jouppiISCA}. Hence, solutions for ultra-efficient DNN processing require a re-thinking of the architecture of computing systems, as well as the adoption of new paradigms at the hardware level. In particular, recent technological breakthroughs in the field analog in-memory computing (AIMC) blurs the distinction between processing and memory can be blurred with custom designed memory, which look beyond conventional von Neumann architectures. With analog in-memory computing (AIMC) certain computations directly take place where the data is located, exploiting device physics and circuit laws \cite{Y2020sebastianNatNano}. In addition to significantly reducing the data movement, AIMC core enables the execution of millions of operations (such as multiply-and-accumulate, MACs)  in parallel, greatly outperforming GPUs and other accelerators.

One approach to exploit in-memory computing for DNNs is to design stand-alone accelerators where multiple AIMC tiles and associated digital logic blocks are interconnected by a suitable communication fabric \cite{Y2016shafieeISCA, Y2021dazziTC}. In such a multi-tile accelerator, weights associated with different neural network layers can be mapped to different AIMC tiles and data can be propagated via tile-to-tile communication. Yet, these designs fall short on supporting (1) a variety of neural networks types, (2) changes in inter-layer connectivity and data-flow, (3) different processing and activation functions, and (4) multiple number formats (e.g. quantization levels), all of which require a high degree of flexibility, difficult to implement solely in hardware. Combining AIMC with general purpose processors can be key for powerful and cost-effective AI platforms, completely avoiding the need for designing generations of dedicated accelerators to cope with the rapid advances in DNN workloads and models.  

One way to address the limitations of standalone AIMC-based accelerators is to add local CPUs~\cite{Y2018feinbergISCA, Y2019ankitASPLOS, Y2020jiaJSSCC, Y2021ottaviAICAS}. The CPU-AIMC interplay nonetheless often is the run-time bottleneck in these systems, as AIMC tiles at competitive technology nodes typically operate on the order of hundreds of nanoseconds~\cite{Y2020xueISSCC, Y2021khaddamaljamehVLSI}. Any communication overhead with the CPU translates into unutilized AIMC tile compute cycles and hence a performance loss. Therefore, it is essential to work towards a tight integration to fully realize the potential of AIMC-based acceleration, including support for a software ecosystem so that systems can be easily configured to implement a wide range of neural networks.% and be compatible with the commonly used ML frameworks.
%Moreover, these aforementioned implementations lack a complete full-stack hardware and software ecosystem, preventing them from being easily adopted by a wide range of networks and by regular DNN programmers using ML frameworks. To unleash the full power of AIMC we need hardware and software to work in a truly coordinated way, by providing easy and programmer-transparent ways of accessing and running workloads in the AIMC tiles. 

In turn, the realization of this vision requires, as a necessary precondition, the availability of flexible, accurate and hardware-validated simulation frameworks. Such frameworks are  crucial to perform the fast exploration of different AIMC integration options from a system-wide viewpoint. They must be capable of targeting entire and complex applications, evaluating the cost/performance of their accelerated and non-accelerated parts (as well as their interplay), taking into account computation and communication.

%\textcolor{red}{Giovanni to Irem: I would not talk about BLADE here (next paragraph), just in the SoA as an example of in-memory computng different from AIMC.} 

%Working towards a seamless integration, BLADE in-cache computing architecture was proposed, where the AIMC paradigm is exploited at the cache level [Sim19GLVLSI, Rio19GLVLSI, Sim20IEEETranComp]. BLADE has been recently implemented in 28 nm CMOS [Sim19DAC], designed from the circuit level, right up to the architecture and system-level to obtain 3x performance and 1.5x better energy efficiency for the inference of DNN on embedded devices. Yet, with a cache size on the order of kilobytes, this technique falls short for implementing even mid-size neural networks.

Towards this end, we herein illustrate a detailed system architecture exploration with a novel full system simulation framework, named ALPINE (or ``Analog In-Memory Acceleration with Tight Processor Integration for Deep Learning"). ALPINE instances feature AIMC crossbars as dedicated components, realised as tightly coupled accelerators, which allow the storage and processing of megabytes of data in constant time complexity. We show that such integration can be realized without hampering the flexibility of CPUs. Moreover, existing hardware and software stacks can be leveraged by only adding lightweight custom extensions to the instruction set in order to govern the AIMC tile. Our contributions are as follows:

\begin{itemize}
    \item We propose a new system simulation framework, extending industry-standard gem5 (with gem5-X extensions), that allows the modeling of systems with AIMC tiles.
    \item Using this simulation framework, we model tightly integrated AIMC tiles, governed by custom instructions extending the ARMv8 64-bit instruction set architecture (ISA) with custom instructions.
    \item We introduce a custom software library, \emph{AIMClib}, to show how AIMC models can be leveraged from the software programmer perspective and streamline the software development process.
    \item We showcase the mapping of different artificial neural network (NN) types (MLP, LSTM, CNN) onto the proposed architecture in both single-core and multi-core cases. We obtain up to 20.5x performance and up to 20.8x energy benefits (in CNNs) with respect to the multi-threaded CPU + SIMD (ARM NEON) implementation.
    \item Using the aforementioned NNs as case studies, we analyze the prevalence of matrix-vector multiplication (MVM) operations as a computational hot-spot in DL workloads, and quantify the application-wide benefits achievable by acceleration via tightly-coupled AIMC tiles, including up to 12.8x/12.5x speedup and energy improvement in MLPs and 9.4x/9.3x speedup and energy improvement in LSTMs.
\end{itemize}

% The remainder of this paper is organized as follows. Section II presents related work on full system simulations and simulations on AIMC-based system. Section III provides background on AIMC acceleration. Sections IV and V present the ALPINE system architectures and their gem5-X implementations. Section VI describes the experimental setup while we present our results with the ALPINE framework for MLP, RNN, and CNN inference in Sections VII, VII, and IX, respectively. Finally, our conclusions are listed in Section X. 

\section{Related work}
\begin{figure*}[t]
    \centering
    \includegraphics[width=16.8cm]{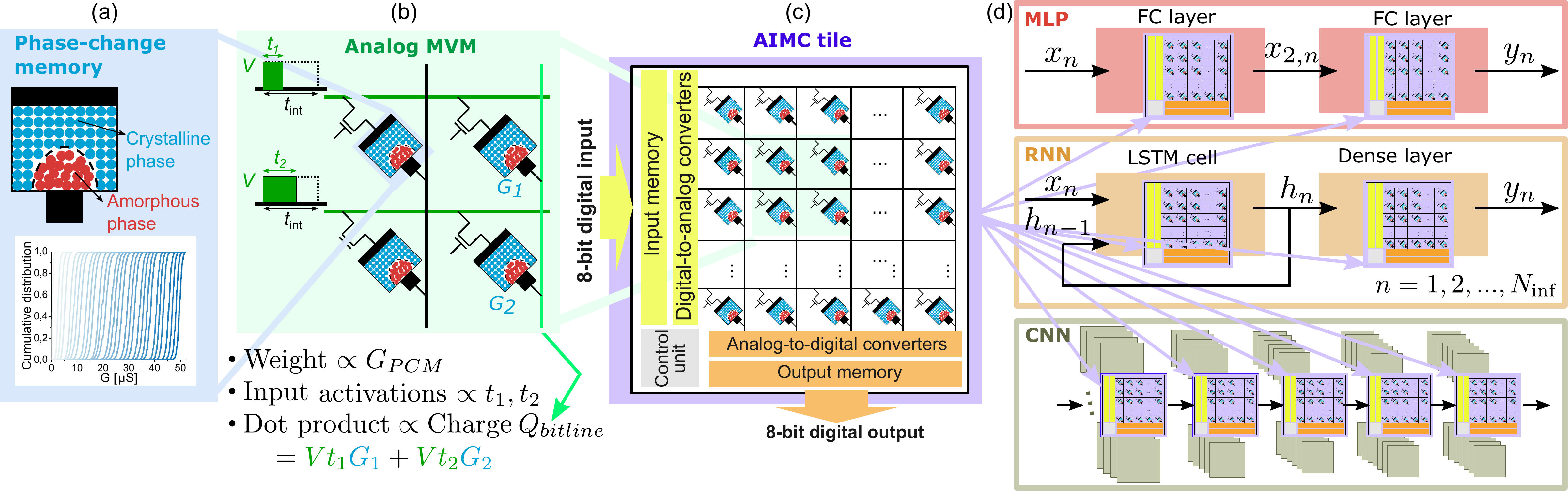}
    \caption{Analog in-memory acceleration for neural network inference. (a) Phase-change memory (PCM) device with true analog storage capability (32 representative levels shown on real hardware experimentally). (b) Matrix-vector multiplication with PCM devices. (c) Analog in-memory computing (AIMC) tile with PCM array, data converters and digital control unit. (d) Weights of MLPs, RNNs and CNNs represented with AIMC tiles.}
    \label{fig:AIMC}
\end{figure*}

\subsection{Full system simulations}

Custom hardware extensions and accelerators can be modeled and simulated at various abstraction levels.  Hardware description language (HDL) and register-transfer level (RTL) simulation frameworks offer the most detailed view of hardware being designed, in addition to a direct path to synthesis and eventually fabrication. %\cite{rocket}.
Yet, the development effort, synthesis/compilation time, and simulation time costs are very high, even with higher levels of hardware abstraction made possible by high level synthesis \cite{hls}.

While the overhead costs of novel hardware development may still be worth the effort for uniquely low-level pieces of hardware (e.g. new functional units), AIMC tiles and similar accelerator technologies are made to work with user-space applications (neural network inference) running on top of an operating system and virtual memory sub-system. The time for simulating full computer hardware and software stacks in conventional HDL-level simulators goes up dramatically and is therefore not scalable when the operating system needs to boot, load user-space programs, and then keep track of OS and low-level processes such as services and interrupts~\cite{hls}.

Due to this complexity, an abstract approach is more suitable for investigating system architectures in a more flexible manner. In broad terms, full system-level simulators simulate crucial components of a computer (CPU cores, functional units, buses, caches, memory, and more) as black boxes in software with tunable latencies and attributes. As a result, simulations can be run significantly faster and with higher degree of flexibility than that of HDL-level simulators. The premiere full system-level simulator in academia and industry is gem5, which is an open-source full system-level simulator that supports multiple ISAs (x86, ARM) and hardware models upon installation, in addition to being regularly updated with support from academia and industry\cite{Y2019qureshiSpringSim}. %\cite{binkert2011gem5}.

Because critical hardware components are simulated in software, full system-level simulators like gem5 must be validated with respect to real hardware in order to produce performance results representative of that one would attain using a HDL-level or RTL simulator. For all of our applications and experiments, we use gem5-X, an extended open-source version of the gem5 simulator. gem5-X is shown to operate with less than 5\%  error on performance statistics in comparison to a real ARM Juno platform (developed by ARM in 2015) \cite{Y2019qureshiSpringSim}.

All of our experiments simulate the full computer architecture stack, including user-space programs running on top of Linux 4.3 and an Ubuntu LTS 16.04 disk image. The simulated full system includes validated hardware models for the CPUs and memory hierarchy.

\subsection{Simulations of AIMC-based systems}

Research efforts investigating AIMC for inference have, for the most part, focused on achieving iso-accuracy with digital systems. Prior work includes iso-accuracy studies for convolutional neural networks (CNNS)~\cite{Y2020joshiNatComm, Y2021kariyappaITED, Y2021fahimiArxiv}, recurrent neural networks (RNNs)~\cite{Y2019tsaiVLSI, Y2021kariyappaITED} and transformers~\cite{Y2021spoonVLSI}. However, very few papers have discussed how to model the potential benefits of AIMC at system level or discuss the integration of AIMC for system acceleration.

Most of the works in this space are illustrated as part of architectural and compilation frameworks studies, and are therefore tightly linked to a hardware/software ecosystem. Among them, Chen et al. introduce an instruction-accurate-only circuit-level simulator for gathering statistics on area, latency, dynamic energy, and leakage power of a synaptic array model \cite{chen2018neurosim}.  Working at a higher level, Kourtis et al. \cite{kourtis2020compiling} introduce a software stack to automate the mapping of CNNs (described in high-level language) onto multi-tile AIMC accelerators. As part of their contribution, the authors showcase a cycle-accurate simulator. Similarly, Ankit et al. \cite{Y2019ankitASPLOS} describes an ISA and compiler dedicated to programming and utilizing a multi-tile AIMC accelerator, and an associated architectural simulator (named PUMASim) to evaluate the energy and performance of compiled applications. The scope of both these approaches is limited to the modelling of the AIMC accelerator alone, neglecting other system components.

Ambrosi et al. \cite{ambrosi2018hardware} propose, as part of their ONNX compilation framework, a set of simulators at three abstraction levels: performance (low-level hardware simulation of single AIMC instructions), functional (component-level simulation of AIMC tiles and associated memories) and system. Our research contribution is most closely related to the latter, which is based on the QEMU emulator %\cite{bellard2005qemu} 
and aims at investigating the execution performance of entire applications. Zheng et al. also use the ONNX framework as the front end for their event-driven cross-level simulation of processing-in-memory accelerators, while also incorporating elements for simulating memory access and interconnects \cite{zheng2022pimulator}.  However, both of these simulators do not take into account the interaction between applications and operating systems, nor does it consider the interplay between AIMC tiles and the rest of the system, including CPUs and, in the case of \cite{ambrosi2018hardware}, the memory hierarchy.

Most closely related to our approach is the paper of Vieira et al., which details a full-system evaluation strategy of AIMC acceleration. As in our case, the authors also base their approach on AIMC-dedicated extensions to the gem5 environment \cite{bdpe}.
%\cite{binkert2011gem5}.
Nonetheless, their approach is limited to modelling the simple case of binary CNNs, and their per-kernel mapping strategy does not scale to the larger and more general applications we tackle in this paper. Moreover, as opposed to our ISA extension enabling multi-core CPU systems with multiple AIMC tiles, their extension only supports a single-core CPU and a single AIMC tile.

Finally, Ottavi et al. \cite{Y2021ottaviAICAS} proposes and synthesizes a heterogeneous architecture where an AIMC tile sits within a cluster of RISC-V processors. They perform a design-space exploration using the bottleneck of MobileNetV2 CNN where point-wise and depth-wise convolutions are placed within the AIMC tile. However, their synthesis approach hinders quick exploration of alternative architectures and model runs.

\section{Background on analog in-memory acceleration}
\subsection{Analog in-memory computing paradigm} 
AIMC offers significant advantages in terms of energy and performance owing to two key properties. First, the computation takes place in the memory and therefore, the expensive data movement can be avoided (addressing the memory read bottleneck). Secondly, the computation can be done in a massively parallel and analog manner by exploiting the physical attributes and state dynamics of memory devices. SRAM-based AIMC approaches are attractive owing to SRAM's technological maturity and scalability to aggressive CMOS nodes~\cite{Y2019biswasJSSCC, Y2021jiaISSCC}. One drawback with this approach is that only a single bit can be stored in an SRAM cell. An alternative is to adopt AIMC based on non-volatile memory technologies, including 2D~\cite{Y2018merrikhBayatTranNNLS} and 3D Flash~\cite{Y2021KimJSSCC}, phase-change memory (PCM)~\cite{Y2021khaddamaljamehVLSI}, and resistive random-access memory (RRAM)~\cite{Y2020xueISSCC}. These technologies offer analog data storage capability, i.e. the ability to achieve a continuum of resistance/conductance values (Fig. \ref{fig:AIMC}(a)). Their non-volatile nature makes them particularly attractive for low-power embedded applications as non-volatile memory-based AIMC tiles consume negligible static power.

In this paper, we focus on PCM for AIMC, which is arguably the most mature technology among the class of resistance-based or memristive memory devices  (Fig. \ref{fig:AIMC}(a)). PCM devices have the potential to scale to nanoscale dimensions and can be integrated in the back-end of a CMOS chip~\cite{Y2021khaddamaljamehVLSI}. PCM-based implementations hence offer high performance densities (TOp/s/mm$^2$), where a pair of PCM devices can represent signed multi-bit weights~\cite{Y2020joshiNatComm}.

MVM operations, which form the bulk of computation for DNN models, can be implemented in a PCM crossbar by representing the elements of a $M$x$N$
matrix  as the conductance values of memory devices, as shown in Fig. \ref{fig:AIMC}(b). Each element of an input vector is translated into the duration of a voltage pulse with fixed amplitude $V$. The voltage pulses are applied simultaneously to $M$ word lines and each memory device contributes to the current flowing through one of the $N$ bit lines, with an amount directly proportional to its conductance $G$ (Ohm’s law). The total current integrated on each of the bit lines over a certain period of time $t_{int}$ is indicative of the result of the dot product between the $M$-element vector and a column of the $M$x$N$ matrix (Kirchoff’s current law). Hence, the multiplication of an $M$x$N$ matrix with an $N$-element vector can be performed in a constant amount of time, (in the range of 10s to 100s of nanoseconds~\cite{Y2021khaddamaljamehVLSI}).

%Tight conductance distributions are desired when programming the weights initially to the crossbar. This can be achieved with feedback-driven close-loop iterative programming via pulse amplitude modulation~\cite{Y2020joshiNatComm}. A diagonal or row-wise parallel programming can be employed to reduce the crossbar programming complexity from $O(N^2)$ to $O(N)$ \cite{Y2021khaddamaljamehVLSI, Y2021narayananVLSI}.

\subsection{Analog in-memory compute tiles}

An AIMC tile contains digital-to-analog and analog-to-digital converters (DACs and ADCs) with dedicated registers and a local controller, alongside the memory crossbar array of unit cells, as presented in~\cite{Y2021khaddamaljamehVLSI}. DACs convert the signed digital input into a voltage pulse; the pulse amplitude is applied either as $V$ or $-V$ according to the sign and the duration of the pulse is proportional to input magnitude.  The dot product over the bit line is converted to a signed digital output via ADCs. Each signed weight is represented with a pair of PCM devices; therefore, a differential bit line current is received by the ADC. The local controller orchestrates the data flow from the data bus into the DAC registers and out of the ADC registers to the data bus. It also activates the MVM operation when input data has arrived into the DAC registers.

In our design, we have a dedicated DAC and ADC for each word line and bit line, respectively. The resolution of DACs and ADCs are signed 8-bits. The input signal is scaled and quantized in digital prior to its transfer to the AIMC tile. This input scaling factor can be arbitrarily selected, preferably fixed to avoid dynamic scaling. Similarly, the ADC quantizes the output of the MVM operation. %Yet, the quantization range cannot be arbitrarily tuned; the ADC gain is typically fixed after an initial calibration step. 

\subsection{Computation precision with analog in-memory computing}
\label{subsec:DL-SoA}

In addition to the quantization introduced by DACs/ADCs, the weights stored in the memory crossbar also have a low precision. Yet, the nature of the precision loss for the weights with analog computing is substantially different from the weight quantization of a digital implementation. The programming and reading of analog weight value are prone to various non-idealities, including noise, temporal conductance fluctuations and temperature-induced variations~\cite{Y2020nandakumarIEDM, Y2021boybatIEDM}.

%In a conventional digital memory, the stored information is read from memory elements individually with sense amplifiers, resulting in relatively low bit error rates. This substantially differs from analog computing where all memory elements on the bit line contribute to the sensed charge. Any deviation from the targeted conductance value  (owing to variability, noise, temporal conductance fluctuations, temperature-induced variations~\cite{Y2020nandakumarIEDM, Y2021boybatIEDM}) or circuit-related effects (such as IR drop) can have severe impact on the model accuracy. 

The scalar multiplication of an analog input with PCM-based weights is shown to be comparable to an implementation with 4-bit fixed-point inputs and weights \cite{Y2018legalloNatElectronics}, and even to an implementation with 8-bit fixed-point inputs and weights with suitable innovations in device design~\cite{Y2018giannopoulosIEDM}. To counter the reduced weight precision, one can employ noise during training, so that the model is more robust when performing inference on AIMC tiles~\cite{Y2020joshiNatComm}. 
An alternative approach is to encode weights using multiple PCM devices~\cite{Y2019tsaiVLSI}. Despite the reduced precision weights, AIMC implementations were shown to address the inference of MLPs~\cite{Y2020nandakumarIEDM, Y2021boybatIEDM}, CNNs~\cite{Y2020joshiNatComm, Y2020nandakumarIEDM, Y2021boybatIEDM}, RNNs~\cite{Y2019tsaiVLSI, Y2020nandakumarIEDM, Y2021boybatIEDM}, and transformers~\cite{Y2021spoonVLSI} with high accuracies.

%The reduced computing precision of AIMC poses challenges to perform DNN inference with an accuracy comparable to digital floating-point implementations. One method to counter the reduced weight precision is to employ noise during training, so that the model is more robust when performing inference on AIMC tiles~\cite{Y2020joshiNatComm}. Other approaches include encoding the model weight using multiple PCM devices~\cite{Y2019tsaiVLSI}, calibrating the MVM results against shifts in statistics~\cite{Y2020joshiNatComm}, and engineering superior devices~\cite{Y2018giannopoulosIEDM}. With the latter, an MVM precision analogous to an 8-bit fixed-point implementation is achieved. Note that 8-bit \textcolor{red}{integer} precision is sufficient to reach state-of-the-art inference accuracies for a wide range of networks \cite{Y2017kapurArxiv, Y2018jacobICCVPR}. 

\section{Architecting ALPINE systems}
% \begin{figure}[t]
%     \centering
%     \includegraphics[width=8.2cm]{figures/3_AIMC-SoC.pdf}
%     \caption{A high-level overview of two AIMC-enabled system architectures. (a) shows the classic loosely-coupled architecture where the CPU accesses an off-chip cluster of AIMC tiles via the I/O bus. (b) configures each CPU core to have direct access to its private AIMC tile in the proposed novel tightly-coupled architecture.}
%     \label{fig:3-AIMC-SoC}
% \end{figure}

\begin{figure}[t]
    \centering
    \includegraphics[width=7.2cm]{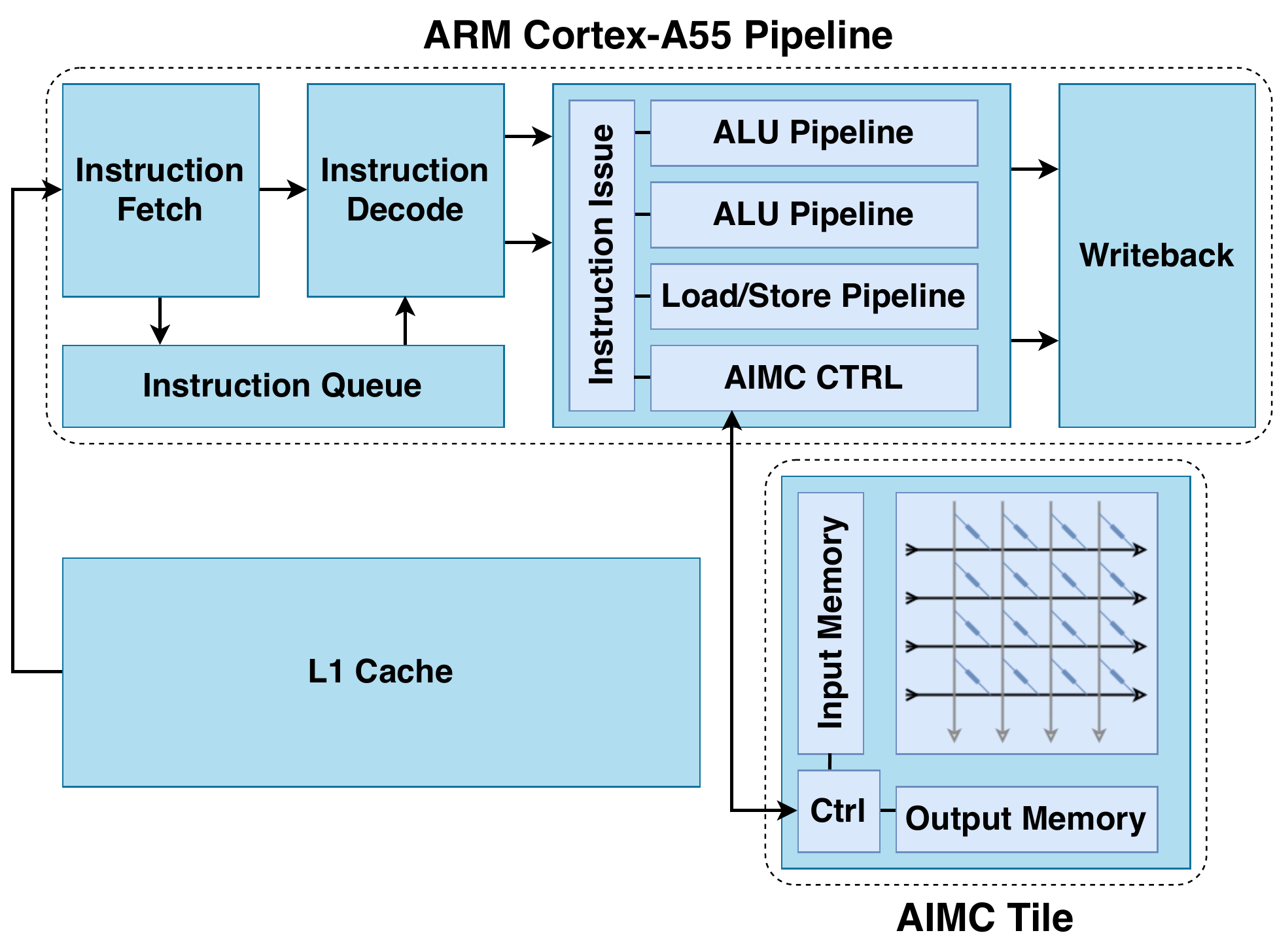}
    \caption{A high-level overview of how a tightly-coupled AIMC tile could interact with an ARM Cortex-A55 pipeline.  A dedicated line from an AIMC control pipeline can send signals and data directly to and from the AIMC tile's control unit, akin to a co-processor.}
    \label{fig:3-AIMC-Tight-Integration}
\end{figure}

%In this section, we give an overview of the potential system architectures that can be supported with the ALPINE framework, discuss the input and output interfaces used to access and interact with AIMC tiles, and then finally present how to interface AIMC tiles from a software designer perspective.

\subsection{AIMC integration strategies}

\textbf{Loosely-coupled AIMC tile-enabled systems:} A high-level overview of a loosely-coupled AIMC-enabled system, can be seen in \cite{Y2018feinbergISCA, Y2020jiaJSSCC}.
%, can be seen in Fig. \ref{fig:3-AIMC-SoC} (a). Here, 
The AIMC tiles are peripheral devices that communicate with the CPU cores via the I/O bus. CPU cores use \textit{load} and \textit{store} instructions to \textit{read} and \textit{write} data to a particular memory-mapped addresses. The AIMC tiles are typically organized as a multi-tile accelerator with a control unit that parses the incoming data so that one or multiple AIMC tiles within the accelerator can be accessed.

While a more common system design approach, loosely-coupling AIMC tiles in this fashion typically leads to a significant communications overhead that can cause the CPU to stall and wait for transactions (empirical data for this assertion is presented in Section 7.B). %This is typically overcome by placing more digital logic on the accelerator containing AIMC tiles, but this strategy comes with numerous drawbacks, including more accelerator energy and power requirements, more fabricated hardware area for the digital logic, and hard constraints on the digital logic capability (i.e., a dedicated ReLU functional unit included in the accelerator model will not be able to handle Sigmoid activation functions or more complex operations that could be required by the network).

\textbf{Tightly-integrated AIMC-enabled systems:} To overcome the communication overhead over the I/O bus, as well as the constraint of flexibility in the loosely-coupled design, we propose a novel tightly-coupled configuration, as seen in Fig. \ref{fig:3-AIMC-Tight-Integration}. Here, the CPU uses an ISA extension to access private AIMC tiles that are unique to each of the CPU cores, without requiring the traversal of the memory hierarchy (Fig. \ref{fig:3-AIMC-Tight-Integration}).

\subsection{Interfacing Tightly-integrated AIMC tiles}

\begin{figure}
    \centering
    \textbf{(a) High-level Data-flow with ISA Extension}\vspace{0.4cm}
    \includegraphics[width=8cm]{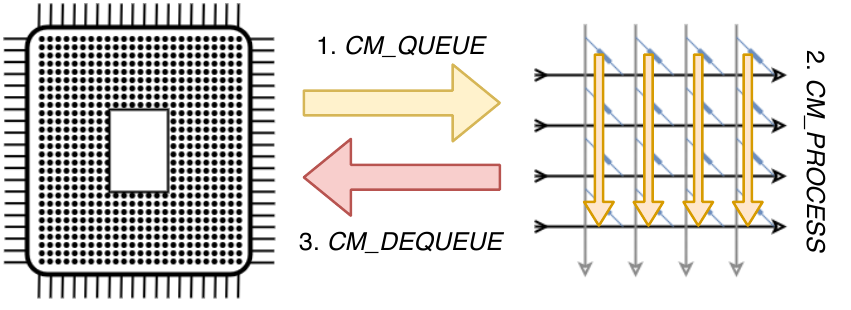}\vspace{0.3cm}
    % \caption{Proposed CM Instruction Definitions}
    \textbf{\\ (b) Proposed CM Instruction Definitions}\vspace{0.4cm}
    \begin{tabular}{|l|p{1.05cm}|c|c|c|c|c|}
        \hline 
        Op &  OpCode &  Rm &  R/W &  Ra &  Rn &  Rd   \\ \hline \hline
        \textit{CM\_QUEUE} & 0x108 & Rm & 1 & Ra & Rn & Rd \\ \hline
        \textit{CM\_DEQUEUE} & 0x108 & Rm & 0 & X & Rn & Rd \\ \hline
        \textit{CM\_PROCESS} & 0x008 & X & 0 & X & X & Rd \\ \hline
        \textit{CM\_INITIALIZE} & 0x208 & Rm & 0 & Ra & Rn & Rd \\ \hline
    \end{tabular}
    
    \caption{(a) Visualization of the AIMC tile data-flow when using the ALPINE ISA extension. (b) The instruction definitions.} %To perform one matrix vector multiplication ($y = M * x$), first \emph{CM\_QUEUE} is called to send $x$ to the AIMC tile's input memory.  Next, \emph{CM\_PROCESS} is called to perform $y = M * x$, which takes the $x$ sitting in the AIMC tile's input memory, performs multiply-and-accumulate operations on each of the crossbar columns, and then stores the output $y$ in the AIMC core's output memory.  Finally, \emph{CM\_DEQUEUE} is used to send $y$ back to the CPU.}
    \label{fig:3-AIMC-ISA-Dataflow}
\end{figure}

In this section, we present the ISA extension for the tight AIMC integration.  We implement the instructions using the previously unused opcodes in the ARMv8 architecture, as listed in Fig. \ref{fig:3-AIMC-ISA-Dataflow} (b). Prior to inference, the AIMC tile is programmed through the \emph{CM\_INITIALIZE} instruction, which writes 8-bit weight values to the indices of the AIMC crossbar. Active during the inference (our region of interest) are the other three instructions, which are utilized as follows.

The program packs 8-bit inputs into a 32-bit argument register. \emph{CM\_QUEUE} is then called to place the packed inputs into the input memory of the AIMC tile. Additional argument registers are used to specify the number of valid inputs packed in the aforementioned argument register, as well as the input memory index. Once all of the inputs are queued into the input memory, \emph{CM\_PROCESS} is called to operate the AIMC tile by converting the values from input memory into analog voltages via DACs, performing the MVM operation with the stationary weights, and storing the results in the output memory after digitizing them with ADCs. Finally, \emph{CM\_DEQUEUE} is called to retrieve packed 8-bit outputs from the AIMC output memory and place them in a destination register. The argument registers specify the number of packed outputs to retrieve and the index. A visualization of this process is in Fig. \ref{fig:3-AIMC-ISA-Dataflow} (a).

% \begin{table}[t]
%     \centering
%     \caption{Proposed CM Instruction Definitions}
%     \begin{tabular}{|l|c|c|c|c|c|c|c|}
%         \hline 
%         Op &  OpCode &  Rm &  R/W &  Ra &  Rn &  Rd   \\ \hline \hline
%         \textit{CM\_QUEUE} & 0010 0001 000 & Rm & 1 & Ra & Rn & Rd \\ \hline
%         \textit{CM\_DEQUEUE} & 0010 0001 000 & Rm & 0 & X & Rn & Rd \\ \hline
%         \textit{CM\_PROCESS} & 0000 0001 000 & X & 0 & X & X & Rd \\ \hline
%         \textit{CM\_INITIALIZE} & 0100 0001 000 & Rm & 0 & Ra & Rn & Rd \\ \hline
%     \end{tabular}
%     \label{tab:3-AIMC-Insts}
% \end{table}

\subsection{Interfacing tightly-integrated AIMC-Enabled systems}

The ISA extensions provide low-level support for tight AIMC integration, however we also developed a higher-level software library, \emph{AIMClib}, to facilitate the development of AIMC-accelerated applications. This library is coded in C and can be used with C or C++ applications. In addition to containing the intrinsics in convenient in-lined wrapper methods (that use the C++ built-in \emph{\_\_asm}), it includes numerous functions and templates, such as the ability to queue/dequeue whole array or vector data structures of AIMC tile input/output memories, tiling matrices at offsets in the crossbar (so that multiple matrices of varying sizes can be placed next to each other), type-casting for tile inputs and outputs between \emph{int8\_t} and higher precision types (e.g. \emph{fp32}), performing activation functions and other digital processing operations on tile outputs, and a checker program that simulates tightly-coupled AIMC tiles in guest software so that programs that utilize \emph{AIMClib} can be debugged on the host machine before engaging the real or simulated hardware.

A sample of C++ pseudo-code using \emph{AIMClib} is presented in Fig. \ref{fig:3-AIMClib-Example-Code}. In addition to its own methods, we have also used \emph{AIMClib} in conjunction with the Eigen C++ library (version 3.8). Eigen optimizes data structure space utilization and access as well as incorporates SIMD vector operations in our test applications~\cite{eigenweb}.

\begin{figure}[t]
    \begin{lstlisting}[language=C++,frame=none]
#include "aimclib.hh"

int main(int argc, char * argv[]) {
    // Mapping weights to the crossbar with
    // x, y offset of 0, 0 using AIMClib.
    int8_t ** weights = ...
    mapMatrix(0, 0, M, N, weights);
    
    for (int i = 0; i < N_INFERENCES; i++) {
        // Queue input array into the AIMC
        // tile input memory using AIMClib.
        queueVector(sizeof(input) /
            sizeof(input[0]), input[i]);
        
        // Perform MVM using AIMClib.
        aimcProcess();
        
        // Dequeue output memory contents into
        // output array using AIMClib.
        dequeueVector(N, output[i]);
    }
    
    return 0;
}
    \end{lstlisting}
    \vskip -1em
    \caption{A sample C++ code for a single fully-connected feed-forward layer programmed onto AIMC tiles with \emph{AIMClib}. At each inference step, the input is loaded and queued into the input memory. This is followed by the MVM via the \emph{aimcProcess} method.  Finally, the contents of the AIMC core output memory are dequeued straight into an output matrix data structure.}
        \vskip -1em
    \label{fig:3-AIMClib-Example-Code}
\end{figure}

\section{System Simulation}
%In this section, we discuss the technical details of the gem5-X integration of AIMC tiles within the ALPINE framework.

%the software implementation work that enables ALPINE extension to model and simulate AIMC tiles.

\subsection{Modeling AIMC Tiles in gem5-X}

% \begin{figure}
%     \centering
%     \includegraphics[width=8.2cm]{figures/4_gem5-AIMC-Object.pdf}
%     \caption{Software view of the AIMC object and AIMC tile object implementation in gem5-X: The AIMC object controls the generation of AIMC tiles (how many, sizes M and N), while the tiles contain parameterizable crossbars and associated input and output memories.  A variety of inputs can be used to control the function of the AIMC tiles which can be forwarded to various outputs (e.g., I/O Bus, CPUs).\textcolor{red}{is this figure essential?}}
%     \label{fig:4-gem5-X-AIMC-Impl}
% \end{figure}

AIMC tiles are implemented in gem5-X using two classes: one class acts as a wrapper with metadata for setting up AIMC tiles, and the second class contains the AIMC tile itself. %A visualization of these classes is depicted in Fig. \ref{fig:4-gem5-X-AIMC-Impl}. 

The wrapper class is designed to encompass both the loose and tight coupling of AIMC tiles. For implementing the loose coupling, the wrapper is defined as a gem5 Peripheral Input/Output (PIO) device where it is accessible by load and store instructions at a specific memory-mapped address. Alternatively, for the tight coupling case, the wrapper can be accessed by dedicated instruction. 

The AIMC tile class contains the actual crossbar and peripheral simulated modeling components. We base our AIMC tile implementation on \cite{Y2021khaddamaljamehVLSI}. Each generated AIMC tile has an input memory array, the memory crossbar, and output memory array. The component dimensions are parameterizable in the wrapper class configuration. 
% Data conversions (ADCs/DACs) are simulated in software when the AIMC tile is interacted with (i.e., a MVM is performed by the AIMC tile).

\subsection{AIMC-Enabled Systems in gem5-X}

After the wrapper object and AIMC tile objects are defined, the next step is to instantiate them on the system and define the proper system interfaces (e.g., functional units and bus controller). This is realized in gem5-X systems through gem5-X's configuration scripts, of which an example is found in Fig. \ref{fig:4-SoC-Config-Code}. By itself, this allows for the system-level implementation of classic loosely-coupled AIMC-enabled systems.% that have permeated prior art by controlling how many AIMC tiles are generated for the entire simulated system and allowing access via bus and DMA interfaces.
To simulate the tightly-coupled AIMC-enabled architectures, we extend the accelerator modeling in \cite{bdpe} such that the custom ARMv8 ISA extension can also interface peripheral I/O (PIO) devices like our wrapper object. For this, we add connections between the ISA extension and PIO device via the system object (e.g., the simulated system that is instantiated on gem5-X's launch). % In our extension, the ISA templates also include the system object which is then subsequently included when the new instruction classes are generated during compilation.  Furthermore, the wrapper class contains a pointer to the system that is set when the system first launches. The result is that when a custom instruction is called in a simulated program running in gem5-X full system mode, the custom instruction accesses a pointer to the system and then the wrapper object -- at that point, the custom instruction can potentially choose which AIMC tile(s) to access and perform AIMC tile operations.  
The latency of the custom instructions is parameterizable, providing modeling flexibility on the AIMC tile. The return value of these instructions (the result held in the destination register) can also be data sent from the AIMC tile to the CPU.  %Furthermore, modeling of the sub-operations of the AIMC tile, including power-gating, analog-to-digital conversion, digital-to-analog conversion, and others, are included in the modeling of the latencies and operation of the custom ISA.

In our implementation for the initial exploration of tightly-coupled AIMC-enabled architectures, we generate one AIMC tile for each CPU core. Note that this is a design choice and the ALPINE framework supports alternative system definitions, including instantiating multiple AIMC tiles per CPU core, a hard-coded number of AIMC tiles, or others. 

%we generate as many AIMC tiles as there are CPU cores in an array, modeling a single AIMC tile for each CPU.  It is worth noting that this configuration is highly parameterizable and configurable: multiple AIMC tiles can be generated per CPU core, a hard-coded number of AIMC tiles can be generated on system boot, CPU cores (via ISA extension) could access other AIMC tiles to make them "public", and so on.  Furthermore, the actual dimensions of each AIMC tile (the size of the input memory, output memory, and crossbar) is also parameterizable.

\begin{figure}[t]
    \begin{lstlisting}[language=Python,frame=none]
# AIMC wrapper class declaration.
class AnalogComputationalMemory(
    BasicPioDevice):
    type = 'AnalogInMemoryComputing'
    cxx_header = "dev/arm/AIMC_Wrapper.hh"

class RealView(Platform):
    type = 'RealView'

    # Instantiation of the AIMC wrapper.
    aimc = AnalogComputationalMemory()
    
    # Connecting the AIMC tile on the bus.
    self.aimc.pio = bus.master
    ...
    \end{lstlisting}
    \caption{gem5-X Configuration Script code example for placing the AIMC wrapper class on the ARM SoC (RealView) platform. The instantiation is done in a similar manner for the loose-coupled and tight-coupled integration.}
    \label{fig:4-SoC-Config-Code}
\end{figure}

\section{Experimental Setup}
\subsection{Target Systems and Power Model}

The system specifications of our gem5-X simulation are listed in Table \ref{Table_Experimental_Setup}-(A). We define two different system configurations to represent different use-cases, namely the \textit{high-power} system such as those in higher-end devices and the high-performance computing domain and the \textit{low-power} system, tailored for the embedded domain and internet-of-things edge devices. We would like to note that a substantial body of work focuses on AIMC tiles in high-power context; yet, the low stand-by power AIMC tile compels an exploration of its integration in low-power contexts as well \cite{Y2021ottaviAICAS, Y2021zhouArxiv}. We use the MinorCPU model in our explorations, which is a 4-stage pipelined CPU with data forwarding and branch prediction.% \cite{binkert2011gem5}.

The power models are shown in Table \ref{Table_Experimental_Setup}-(B).  Our core and cache power model is based on a \unit[28]{nm} bulk system with an ARM Cortex A53 core \cite{Y2019qureshiSpringSim}, while our DRAM power model is based on \cite{lee2018leveraging}. The core and cache power model is comprised of active and WFM (wait for memory) CPU core energy per cycle, as well as energy/power for the last-level cache (LLC).

\begin{table}
\caption{Experimental Setup}\label{Table_Experimental_Setup}
\centering
\begin{tabular}{l}
\textbf{(A) gem5-X FS Mode System Configurations} \\ \\
{\begin{tabular}{ | p{4.2cm} || P{1.4cm} | P{1.4cm} | }
        \hline
        System & Low-Power & High-Power \\ \hline
        CPU Core Model & \multicolumn{2}{ c | }{Minor (In-Order) CPU} \\ \hline
        Number of CPU Cores & \multicolumn{2}{ c | }{8} \\ \hline
        ISA & \multicolumn{2}{c | }{ARMv8 (AArch64)} \\ \hline
        CPU Core Frequencies & 0.8GHz & 2.3GHz \\ \hline
        Supply voltage VDD & \unit[0.75]{V} & \unit[1.3]{V}  \\ \hline
        L1 Data/Instruction Cache Size & 32kB & 64kB \\ \hline
        LLC Cache Sizes & 512kB & 1MB \\ \hline
        Memory Model & \multicolumn{2}{c | }{8GB DDR4 @ 2400MHz} \\ \hline
        Memory Bus Width & \multicolumn{2}{c | }{16b} \\ \hline
        Memory Bus Frontend Latency & \multicolumn{2}{c | }{3 cycles} \\ \hline
        Memory Bus Forward, Response, and Snoop Latencies & \multicolumn{2}{c | }{4 cycles} \\ \hline
    \end{tabular}}
\\ \\ \\ \textbf{(B) System Energy and Power Figures} \\ \\
 {\begin{tabular}{ | p{3.8cm} || P{1.6cm} | P{1.6cm} | }
        \hline 
        System & Low-power & High-power \\ \hline
        Idle Core Energy (pJ/Cycle) & 10.72 & 126.03 \\ \hline
        WFM Core Energy (pJ/Cycle) & 46.04 & 638.99 \\ \hline
        Active Core Energy (pJ/Cycle) & 60.92 & 845.39 \\ \hline
        Mem Controller + IO Power (W) & 3.03 & 5.82 \\ \hline
        LLC Leakage (mW/256kB) & 271.62 & 874.08 \\ \hline
        LLC Read Energy (pJ/Byte) & 1.81 & 5.60 \\ \hline
        LLC Write Energy (pJ/Byte) & 1.63 & 5.02 \\ \hline
        DRAM Energy (pJ/Access) & \multicolumn{2}{ | P{3.2cm} | }{120.0} \\ \hline
    \end{tabular}}
    \\ \\ \\ \textbf{(C) AIMC Tile Performance and Energy Figures} \\ \\
    { \begin{tabular}{| p{5cm} || p{2.35cm} | }
        \hline
        %Operation latency of \textit{CM\_QUEUE}, \textit{CM\_DEQUEUE} & \unit[1]{ns} \\ \hline
        AIMC tile crossbar size & $M$ rows, $N$ columns \\ \hline
        AIMC tile latency & \unit[100]{ns} \\ \hline
        AIMC tile input/output data throughput & \unit[4]{GB/s} \\ \hline
        Input/output memory SRAM capacity & $M$B/$N$B \\ \hline
        Supply voltage VDD (analog) & \unit[0.8]{V}, \unit[1.2]{V}  \\ \hline
        Supply voltage VDD (digital) & \unit[0.8]{V} \\ \hline
        256x256 AIMC tile MVM energy efficiency  & 12.8 TOp/s/W \textsuperscript{*} \\ \hline
    \end{tabular}}
\end{tabular}
 { \begin{tabular}{| p{4.7cm} || p{2.6cm} | }
\multicolumn{2}{l}{
  \begin{minipage}{8cm}
    \footnotesize{\vspace{0.2cm} * The AIMC tile MVM energy is re-calculated for varying tile sizes, considering the crossbar array size as well as data converters.} 
  \end{minipage}
  }
  \end{tabular}}
  \vskip -1.0em
\end{table}

We calculate the total energy of the system using the gem5-X statistics. The generated statistics include total CPU cycles, simulated time, and cache/memory accesses for each experiment run. The full system energy is then the sum of the energies for the core, cache, and DRAM components.

\subsection{AIMC Setup and Modeling}

% \begin{table}[t!]
%     \centering
%     \caption{AIMC core performance and energy figures}
%     \begin{tabular}{| p{3.8cm} || p{3cm} | }
%         \hline
%         %Operation latency of \textit{CM\_QUEUE}, \textit{CM\_DEQUEUE} & \unit[1]{ns} \\ \hline
%         AIMC tile latency & \unit[100]{ns} \\ \hline
%         AIMC tile IO bandwidth & \unit[4]{GB/s} \\ \hline
%         Supply voltage VDD (analog) & \unit[0.8]{V}, \unit[1.2]{V}  \\ \hline
%         Supply voltage VDD (digital) & \unit[0.8]{V} \\ \hline
%         AIMC tile energy efficiency  & 20 TOp/s/W \\ \hline
%     \end{tabular}
%     \label{tab:cm-core-params}
% \end{table}

 Table \ref{Table_Experimental_Setup}-(C) reports the performance and energy metrics of the AIMC tile estimated from hardware measurements and chip designs in \unit[14]{nm} technology node \cite{Y2021khaddamaljamehVLSI, Y2020nandakumarFN}. For compatibility with the core and cache model in \unit[28]{nm} node, we upscale the AIMC tile power estimates with a scaling factor of 5.3x for the high-power system and 2x for the low-power system. These factors are calculated following the classical scaling theory under constant frequency with the formulation $\alpha\beta^2$, where $\alpha$ denotes the dimensional scaling and $\beta$ is the voltage scaling factor~\cite{Y2019shahidiIEEEAccess}.
 
 Note that it is not a straightforward exercise to provide a simple power scaling factor for a mixed-signal design, such as our AIMC tile. One reason for this is the availability of different technology types or processes (e.g., high-performance, low-power, planar, FinFET) with specifications changing across foundries~\cite{Y2017stillmakerIntegration}. Secondly, digital and analog circuits follow different power scaling trends with the technology node~\cite{Y2007kingetBipolar}. Given that analog circuits scale less aggressively in comparison to its digital counterpart, our scaling  represents a rather conservative estimate in this respect. We assume that the AIMC tile performance remains constant between the two technology nodes.

\begin{figure*}[!ht]
    \centering
    \includegraphics[width=14.0cm]{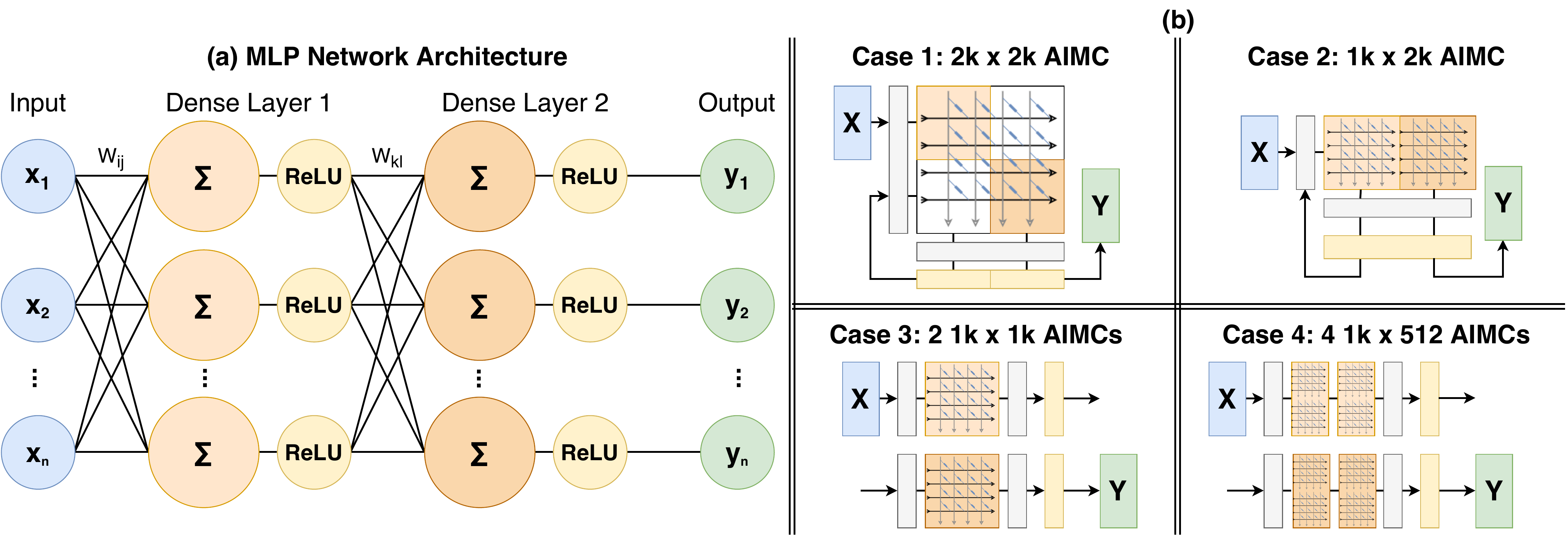}
    \caption{(a) Multi-layer perceptron architecture with two dense (fully-connected) layers (1024, 1024) and ReLU activation. (b) Cases 1 through 4 shows the mappings of the full-connected layers to a variety of AIMC-based configurations.}
    \vskip -1.0em
    \label{fig:6-mlp-cases}
\end{figure*}
\begin{figure*}
    \centering
    \includegraphics[width=16.8cm]{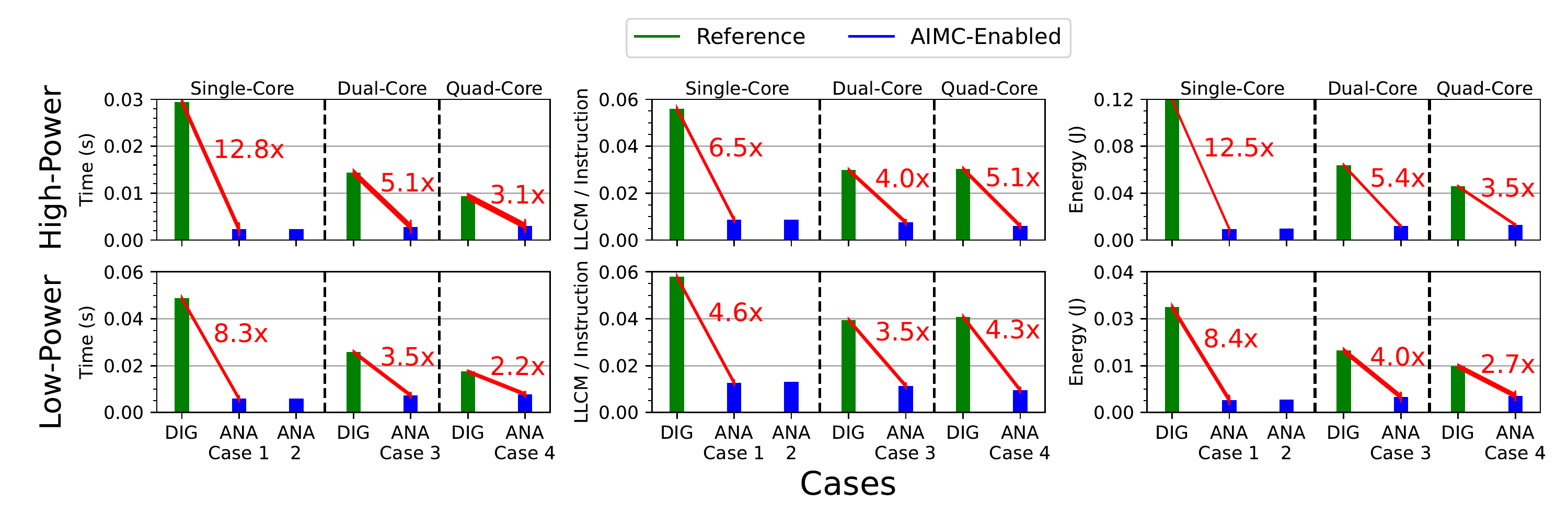}
    \vskip -1.0em
    \caption{Aggregate results for multi-layer perceptron experiments. From left to right, each column contains total time, memory intensity, and energy results for the High-Power system (top row) and Low-Power system (bottom row) configurations. "ANA" refers to analog AIMC-enabled application mappings with implementation numbers corresponding to those in Fig. \ref{fig:6-mlp-cases}, while "DIG" refers to a digital reference or CPU-only implementation.  Results are also grouped by the number of CPU cores utilized (1, 2, or 4).}
    \vskip -1.0em
    \label{fig:6-mlp-aggregate-results}
\end{figure*}

\subsection{Exploration Studies Overview}

To showcase ALPINE's abilities and explore the benefits of tightly-coupling AIMC tiles in systems, we built and optimized a wide variety of neural network models using the Eigen C++ library and AIMClib. For multi-core experiments, we use libpthread to pipeline layers across cores, and implement ping-pong buffering to prevent input/output blocking \cite{eigenweb}.% \cite{libpthread}.  

With these tools in hand, we then perform three neural network explorations.  First, we consider a Multi-Layer Perceptron (MLP) to gauge the performance and energy benefits of a neural network that almost solely relies on matrix-vector multiplication (MVM) operations. Then we consider an alternative neural network architecture, the Long Short-Term Memory (LSTM), that has increased computational requirements outside the MVM. Finally, we look at the fully-pipelined implementation of a Convolutional Neural Network (CNN) and explore how the proposed system behaves in the presence of large number of MVMs in conjunction with very intense memory access patterns.  We further breakdown these applications into multiple cases with different AIMC tile and CPU core mappings.  All of these AIMC tile-enabled neural networks and their implementations are compared against a digital-only, SIMD-enabled, reference application, which employs the same aforementioned optimizations (Eigen Library integration, pthreads, ping-pong buffers).  Furthermore, for more equitable performance comparisons, we use similar precision across all applications (\emph{int8\_t} with \emph{fp32} accumulation where floating point operations apply, such as in sigmoid and softmax operations).

In general, inference-to-inference we notice a deviation in performance results and system metrics of less than 4\%.  Thus, to save on simulation time, we only perform 10 inferences for each of the cases in the MLP and LSTM neural networks.  We further reduce the number of inferences in the CNNs to 3 due to the larger network requiring more simulation time.

%For the MLP and LSTM applications we consider multiple mappings of the neural network architecture that explore single-core and multi-core system benefits. In the CNN application we explore multi-core implementations that utilize all CPU cores available on the simulated SoC.

\section{Exploration One: Multi-Layer Perceptrons}
\subsection{The multi-layer perceptron architecture and cases}

In this first case study, we focus on a two-layer MLP neural network (1024, 1024) with ReLU activation functions (Fig. \ref{fig:6-mlp-cases} (a)). 
%While this acts as a simple example, it is worth noting that MLP-based implementations are becoming more prevalent through recommender systems~\cite{Y2019naumovArxiv} and vision~\cite{Y2021tolstikhinArxiv}. 
%Although MLPs predate other neural network model classes, Google reported that it is still a significant part of their workloads~\cite{Y2021jouppiISCA}.

%The MVMs form the bulk of the MLP inference. The large feed-forward layers are particularly amenable for AIMC acceleration. Therefore, this 

%Because the activation functions are computationally simple, the main component of the MLP run-time (without the use of the AIMC cores) is the matrix-vector multiplication (MVM) via the multiply-and-accumulate (MAC) operations in the two fully-connected layers.  The AIMC tiles specifically speed up these operations, so we can gauge the maximum performance benefits attainable by leveraging the AIMC cores.

We create four different analog MLP implementations where vary the size and number of AIMC tiles, as shown in Fig. \ref{fig:6-mlp-cases} (b). More specifically, Cases 1 and 2 are single-CPU core architectures which use one large AIMC tile. Case 3 is a dual-CPU core architecture with one fully-connected layer assigned to each CPU core. Each AIMC tile is also smaller in capacity relative to the previous cases. Finally, Case 4 is a quad-CPU core architecture with one fully-connected layer's computation being split between two CPU cores. The two CPU cores of the first layer sync their outputs via mutexes before letting the second layer start its processing. Additionally, we compare these implementations with a conventional CPU-only and SIMD-enabled architecture.

\subsection{Single-Core Results and Analysis}

\begin{figure}
    \centering
    \includegraphics[width=8.2cm]{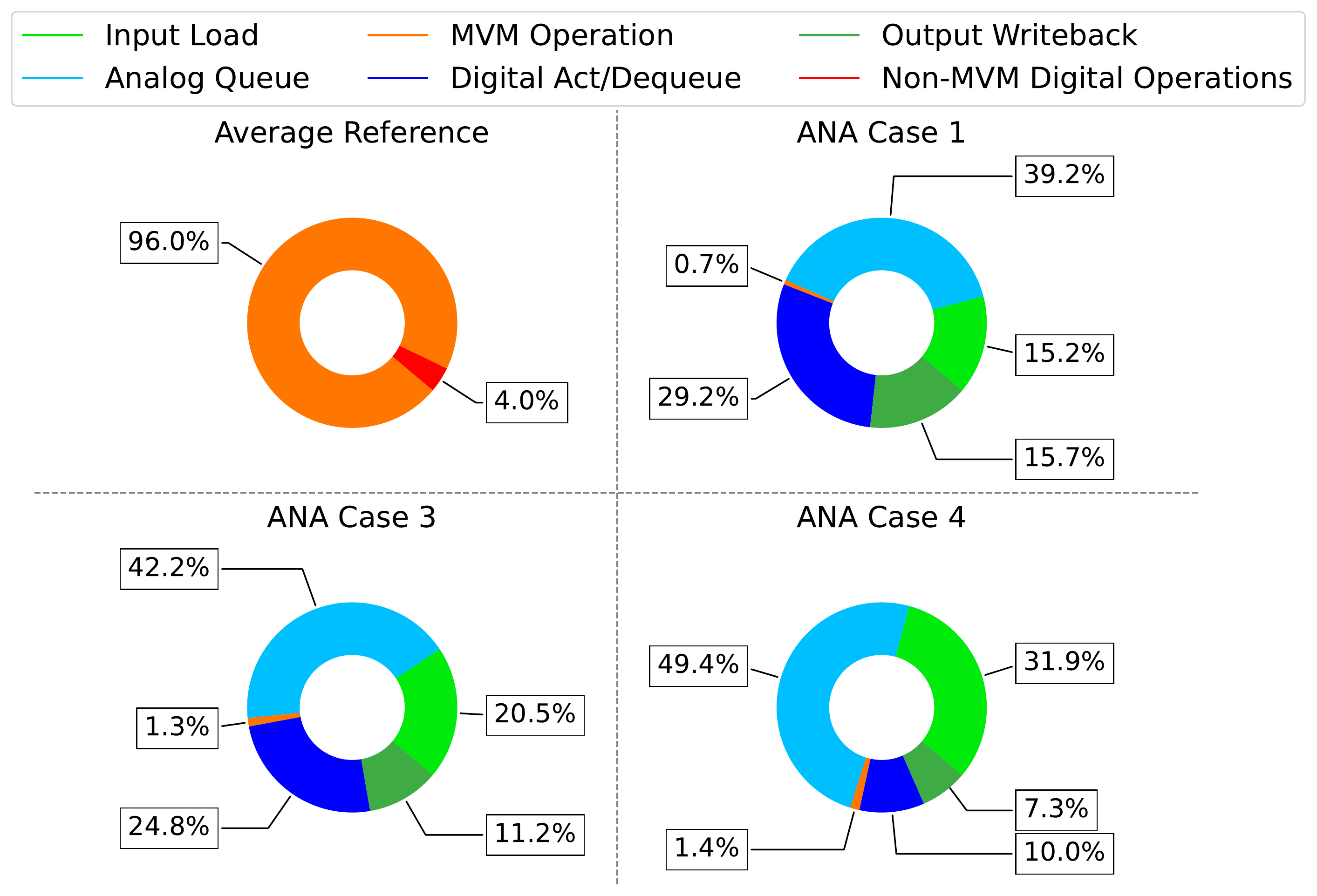}
    \caption{The run-time analysis of MLP cases. Run-time percentage of each sub-ROI division in MLP cases averaged across high-power and low-power systems. The \textit{average reference} is obtained across single-, dual-, and quad-core digital cases. \textit{Non-MVM Digital Operations} refers to the combined run times of the input load, digital activation, and output writeback.  The standard deviation of the timing distribution is less than 1.7\%, 1.2\%, 2.3\%, and 1.5\% for the average reference, Case 1, Case 3, and Case 4, respectively.  The time distribution of cases 1 and 2 are similar and hence, Case 2 is omitted for brevity.}
    \vskip -1.0em
    \label{fig:6-mlp-time-distr}
\end{figure}

Aggregate results for all of our MLP experiments are shown in Fig. \ref{fig:6-mlp-aggregate-results}. We focus on the run-time, energy footprint and memory intensity (quantified as last-level cache misses per instruction, or LLCMPI). The latter provides important insights onto the data movement overhead, which is a significant driver for AIMC-based acceleration. 

%Moreover, we have performed an in-depth run-time analysis of the different implementations, as shown in Fig. \ref{fig:6-mlp-time-distr}. We observe that the MVMs is the main bottleneck for all the digital implementations, with ?90\% of the run time. Meanwhile, in AIMC-tile enabled cases, the MVM accounts for solely < 1.5\% of the run-time.

In all implementations, AIMC tiles provide significant benefits in terms of latency and energy in comparison to the CPU-only runs. Looking at the single-CPU implementations, we observe that Case 1 out-performs Case 2 by a slight margin in terms of latency and energy. While a similar amount of queuing and dequeuing takes place in both implementations, the \textit{CM\_PROCESS} instruction needs to be called twice as much in order to perform the same number of inferences in Case 2. Yet, this does not translate into a 2x run-time and energy-overhead, because the MVMs account for only a small fraction of the total run time, as shown in Fig. \ref{fig:6-mlp-time-distr}. Because of the magnitude of the analog queue/dequeue times on total run time shown in this figure, we show that it is critical to provide a sufficiently large queue/dequeue bandwidth to maximize the benefits from AIMC acceleration, as also discussed in \cite{Y2021ottaviAICAS}.

In addition to the experiments showcased in Figure \ref{fig:6-mlp-cases}, we also tested a "loosely-coupled" AIMC tile-enabled system that places two pipelined AIMC tiles with dedicated ReLU activation units next to each other in an off-chip accelerator.  A single CPU core handles data transactions like sending inputs and receiving outputs from the MLP, which is mapped to the two AIMC tiles.  While this configuration attains 4.1x speedup over the single-core digital reference MLP, it has up to a 3.1x slowdown when compared to the tightly-coupled AIMC-enabled system.%, thus proving the merit of looking at the less communication-constrained tightly-coupled systems.

\subsection{Multi-Core Results and Analysis}
The latency and energy for the multi-core Cases 3 and 4 are displayed in Fig. \ref{fig:6-mlp-aggregate-results}, as well as their time distribution in Fig. \ref{fig:6-mlp-time-distr}. What is immediately apparent with respect to the single-CPU core implementations is that adding multiple CPU cores does not equate to more performance gains. In fact, the performance and energy of the system worsens with increasing number of CPU cores: the single CPU-core Case 1 has approximately 20\% and 30\% better run time over the dual CPU-core Case 3 and quad CPU-core Case 4, respectively.

This slowdown is mainly attributed to the communication overhead of sending inputs and activations across CPU cores. As the number of CPU cores increase, we observe that the run-time associated with \textit{input load} and \textit{analog queue} make up a larger portion of the overall run-time (Fig. \ref{fig:6-mlp-time-distr}). More specifically, in the dual-CPU core implementation Case 3, the overhead of communicating the output from the first layer to the input of the second layer significantly increases the total run time of the application. The overhead is then compounded in the quad-CPU core implementation Case 4 where the input from memory as well as the intermediate activations must be sent to two different CPU cores. The synchronization overhead associated with mutexes of both layers aggravate this as well.  Finally, it is worth noting that the performance impact of the CM\_PROCESS operation becomes negligible in the AIMC tile-enabled algorithms, so even estimates of the latency increased 10x are observed to have minimal impact on the performance results.

Interestingly, the memory intensity remains almost constant across all implementations (Cases 1 through 4), suggesting that memory access is no longer a significant bottleneck. Instead, we conclude that for an application whose run-time is dominated by \textit{input load} and \textit{analog queue}, the overhead of sharing data is no longer negligible and should be treated as the primary bottleneck to gains in run time and energy.

Finally, across all cases, the low-power system exhibits lower performance gains in comparison to the high-power system. This is primarily due to the smaller L1 cache size of the low-power system configuration. A smaller cache requires more requests (and therefore endures more delay) to the L2 cache and memory, which is reflected in the slightly higher memory intensity metric relative to the high-power system.

\subsection{MLP Computational Complexity}
In this section, we present a computational complexity analysis for the CPU-only and AIMC-based implementations. We will assume that the limited cache size and cache trashing, as well as the SIMD operations, do not impact the computational complexity. 
For the CPU-only run, each fully-connected layer's MVM operation has a quadratic complexity ($O(n^2)$), while the corresponding activation function (ReLU), as well as loading initial inputs (\textit{input load}) from memory and storing outputs (\textit{output writeback}), has a linear complexity ($O(n)$). Given that the MLP experiments run for $N_{inf}$ inferences, the total complexity of the MLP run can be formulated as $N_{inf}*(2O(n^2) + 4O(n)) \approx O(N_{inf}n^2)$.

With the introduction of the AIMC tiles however, the complexity of MVM operations reduces to constant time ($O(1)$), assuming that the entirety of the weights of the fully-connected layer can fit in the AIMC tile. Therefore the total computational intensity reduces to $N_{inf}*(2O(1) + 6O(n)) \approx O(N_{inf}n)$ after including complexities for analog queueing, shifting the dominating run-time factor to the linear operations (queuing inputs/dequeuing outputs; Fig. \ref{fig:6-mlp-time-distr}).

\subsection{MLP Memory Requirements}

In this section, we analyze the memory footprint for the CPU-only and AIMC-based implementations. In the CPU-only implementation, the weights of the  fully-connected layers must be loaded from the main memory into L1/L2 caches at every inference. Yet, this is not true in the analog implementations.

Let us define the \emph{working set} as the required amount of data memory per inference. In the CPU-only implementation, this includes the weights of the fully connected layers ($2W$), the inputs loaded from memory ($x$), the intermediary activations ($l_1$), and the final outputs stored to memory ($y$). For our implementation with 8-bit weights, inputs, activations and outputs, the working set size is $2W + x + l_1 + y = 2 * n^2 + 3n \approx 2.1$MB for $n = 1024$. For all of our experimental configurations, this working set size exceeds that of both the private L1 caches and the shared L2 cache, meaning elements of the fully-connected layers and the input/output must be thrashed (swapped in and out of the caches, as well as potentially main memory), leading to both worse memory performance and worse total run time.

In contrast, the AIMC-enabled MLP keeps all of the weights of the fully-connected layers stationary inside the AIMC tiles.  After the one-time cost for programming the weights in our MLP, the weights are never reprogrammed, and therefore, can effectively be removed from the working set. In this case, the working set size can be formulated as $x + l_1 + y = 3n \approx 3$kB, which fits comfortably in L1 private caches for both of our test system configurations. This leads to lower memory intensity, less cache thrashing, and thus improved overall performance.

The reduction in computation and throughput requirements resulting from the introduction of the AIMC tiles in the single-CPU core cases 1 and 2 is related to the ones obtained with multi-threading (cases 3 and 4), with the caveat that the computational complexity goes down by the number of hardware threads and space complexity is distributed across the CPU cores. However, even though the space complexities are reduced, the impact of the linear computational complexities of \textit{input load} and \textit{analog queue} are increased by the emerging core-to-core communications bottleneck in the multi-CPU core applications (cases 3 and 4).

We therefore reiterate that when AIMC tiles are introduced to neural networks with very small digital operation requirements (such as only ReLU activation functions), that core-to-core communications overhead should be minimized by using fewer CPU cores and AIMC tiles possible.  By distributing simple digital computation across numerous CPU cores, core-to-core communication emerges as the new bottleneck and can hinder, rather than help, the performance of multi-core-enabled neural networks.% Furthermore, greater performance gains can be achieved by localizing digital logic requirements such that more linear operations are performed by the same CPU.

\section{Exploration Two: Long Short-Term Memory}
\subsection{LSTM Architecture}

% Moved this figure here to coax it into being on the correct page... 
\begin{figure*}[t]
    \centering
    \includegraphics[width=14.0cm]{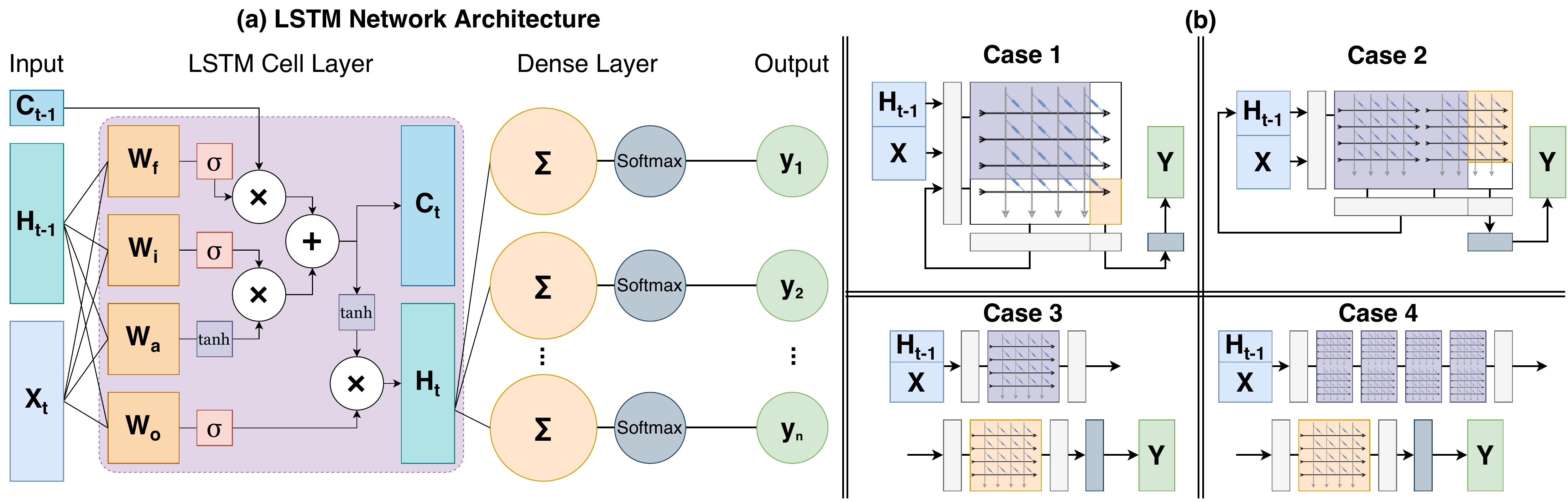}
    \caption{(a) The neural network diagram for the LSTM modeled in our AIMC-enabled test programs.  It is a 2-layer LSTM with one LSTM hidden cell layer and one dense (fully-connected) layer with various, more compute-intensive, activation functions. Note that the AIMC sizes are variable depending on the size of the hidden cell layer and that all activation functions are performed in the CPU cores. (b) Cases 1 through 4 shows the mappings of the layers to a variety of AIMC-based configurations. }
    \label{fig:7-lstm-cases}
\end{figure*}

\begin{table}
\caption{LSTM experiment setup}\label{tab:7-lstm-params}
\centering
\begin{tabular}{l}
\textbf{(A) LSTM Neural Network Parameters} \\ \\
    {\begin{tabular}{|c|c|c|c|}
        \hline 
        Input ($x$) &  Hidden Layer ($n_h$) & Output ($y$) & Total Parameters \\ \hline \hline
        50 & 256 & 50 & 377.3k \\ \hline
        50 & 512 & 50 & 1.28M \\ \hline
        50 & 750 & 50 & 2.6M \\ \hline
    \end{tabular}}
\\ \\ \\ \textbf{(B) LSTM AIMC Tile Dimensions} \\ \\
   { \begin{tabular}{|c||c|c|c|c|}
        \hline 
        $n_h$ & Case 1 & Case 2 & Case 3 & Case 4 \\ \hline \hline
        256 & 612 x 1074 & 356 x 1074 & 356 x 1024 & 356 x 256 \\ \hline
        512 & 1124 x 2098 & 612 x 2098 & 612 x 2048 & 612 x 512 \\ \hline
        750 & 1600 x 3050 & 850 x 3050 & 850 x 3000 & 850 x 750 \\ \hline
    \end{tabular}}
\end{tabular}
\end{table}

%In the previous section, we examined how the AIMC tiles can benefit applications where the primary overhead is the MVM operation.  However, another perspective one can take of tightly-integrated AIMC tiles is that they are a co-processor with respect to the CPU, which can help to decrease the communications latency of interacting with digital logic. Furthermore, tightly-integrated AIMC tiles allow us to leverage preexisting hardware for operations that would be expensive (either with respect to area or communications latency) to perform in a loosely-coupled configuration.  And thus f

In our second exploration, we look at recurrent neural networks (RNNs) in the form of a Long-short term memory LSTM targeting character recognition using the Penn Treebank (PTB) data set \cite{Y1993marcusComput}. The LSTM has one cell (hidden) layer and one fully-connected layer, as presented in Fig. \ref{fig:7-lstm-cases} (a). In comparison to the MLP, the LSTM features more computationally heavy digital operations (sigmoid, tanh, softmax). Moreover, the data flow bears differences owing to the recurrent connection of the LSTM cell. %Therefore, LSTM allow for the evaluation of performance benefits of leveraging AIMC tiles when digital operations play a larger role in overall execution.
Figure \ref{fig:7-lstm-cases} (b) shows the different simulated cases. Cases 1 and 2 are single-CPU core cases that use larger AIMC cores. Case 3 is a dual-CPU core case with the cell layer assigned to the first CPU core, and the dense layer assigned to the second CPU core. Finally, Case 4 is a quin-CPU core case with the cell layer's computation split across the first four CPU cores, and the dense layer assigned to the last CPU core. Here, the four CPU cores associated with the LSTM cell sync their outputs via mutexes before the second layer starts its MVM operation. 

In this exploration, we focus on LSTM instances sharing the same architecture but with different layer sizes. The dimensions of the layers in these networks, as well as the corresponding AIMC tile sizes, are listed in Table \ref{tab:7-lstm-params}. %To evaluate the decreased computational complexity hypothesized and elaborated upon in the first case study, the dimension of the cell layer, $n_h$, is also varied and we examine the subsequent changes in performance results.  Like before, we assume the entirety of the layer can fit inside the AIMC tiles.  
To reduce CPU core-to-core communication as much as possible in case 4, the LSTM cell layer is mapped to AIMC tiles such that instead of the gates being distributed to different AIMC tiles, they are sliced so that element-wise operations can be performed by reading four consecutive columns \cite{Y2020nandakumarFN}.

We would like to note that the largest LSTM architecture ($n_h = 750$) is shown to experimentally yield high accuracy when implemented on real PCM prototype hardware chip \cite{Y2021boybatIEDM}.

\subsection{Single Core Results and Analysis}

\begin{figure*}
    \centering
    \includegraphics[width=16cm]{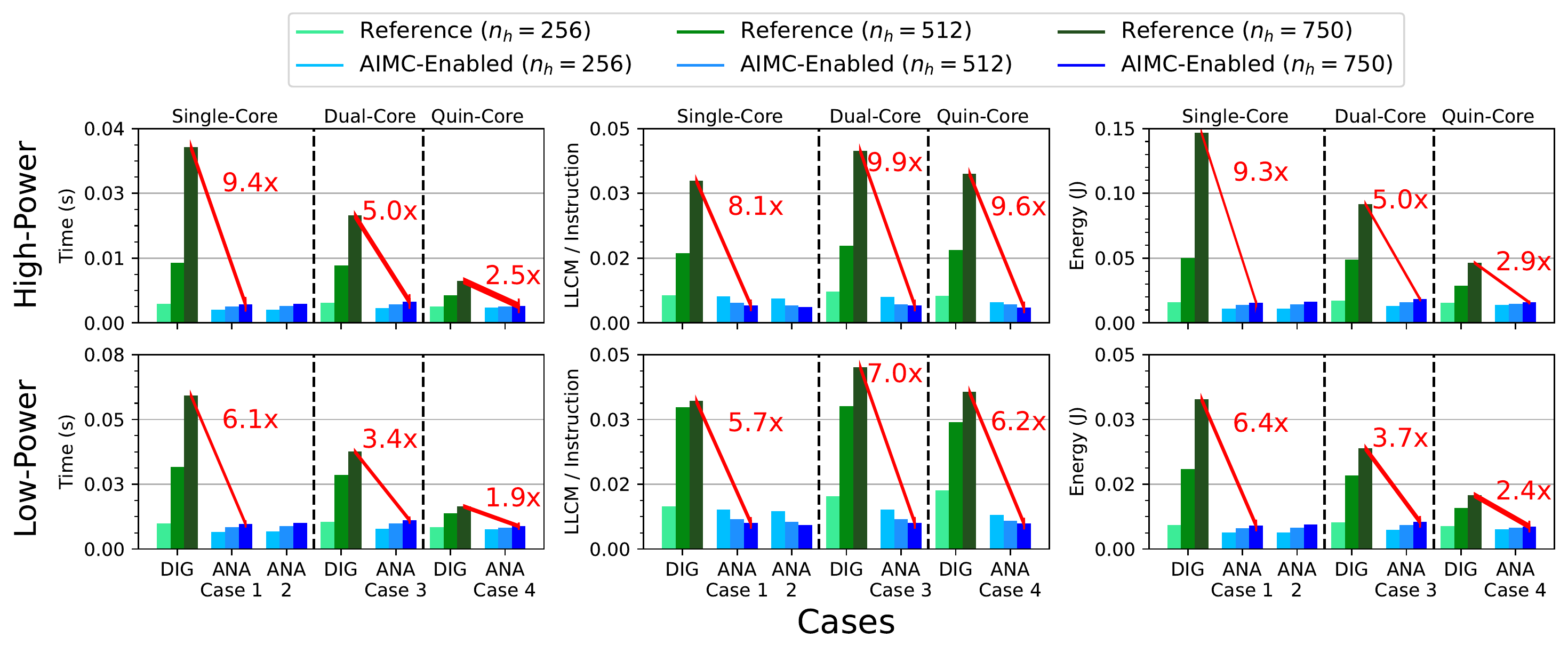}
    \vskip -1.0em
    \caption{Aggregate results for all LSTM experiments.  From left to right, each column contains total time, memory intensity, and energy results for the high-power system (top row) and low-power system (bottom row) configurations.  "DIG" refers to digital reference applications while "ANA" refers to a specific analog, AIMC-enabled application case, which correspond to those in Figure \ref{fig:7-lstm-cases}.  Results are grouped by the number of CPU cores utilized, and from left to right, each grouped column refers to a different $n_h$ parameter which affects the total size of the network.  The darker bars refer to a larger $n_h$.}
    \label{fig:7-lstm-aggregate-results}
\end{figure*}

\begin{figure}
    \centering
    \vskip -1.0em
    \includegraphics[width=9cm]{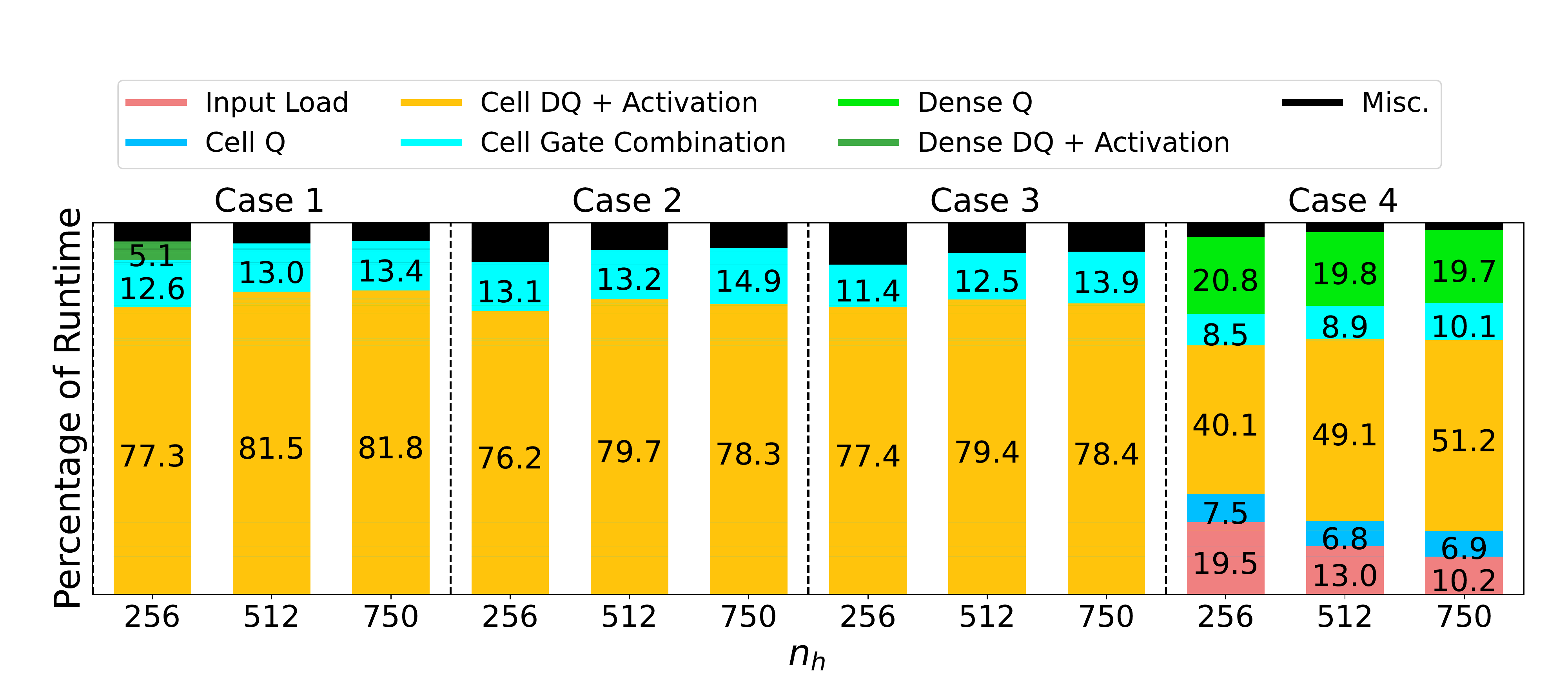}
    \caption{Bar graphs showing the percentage of the ROI run time going towards sub-ROIs for the analog cases 1 through 4 run on the high-power system configuration for all $n_h$ values.  The trends of the high-power system sub-ROI run times are followed by the same cases in the low-power system configuration.  The digital cases (single-, dual-, and quin-CPU core) see 87.8\% to 97.9\% of their total ROI run time dedicated to the digital MVM operations with activation functions in the cell layer. The "Misc." run time is comprised of all other sub-ROIs per experiment that comprise of less than 5\% of the total ROI run time.}
    \label{fig:7-lstm-subroi}
\end{figure}

% Moved this here to coax it into being on the correct page...
\begin{figure}[t]
    \centering
    \textbf{(a) CNN Architecture}\vspace{0.4cm}
    \includegraphics[height=2.7cm]{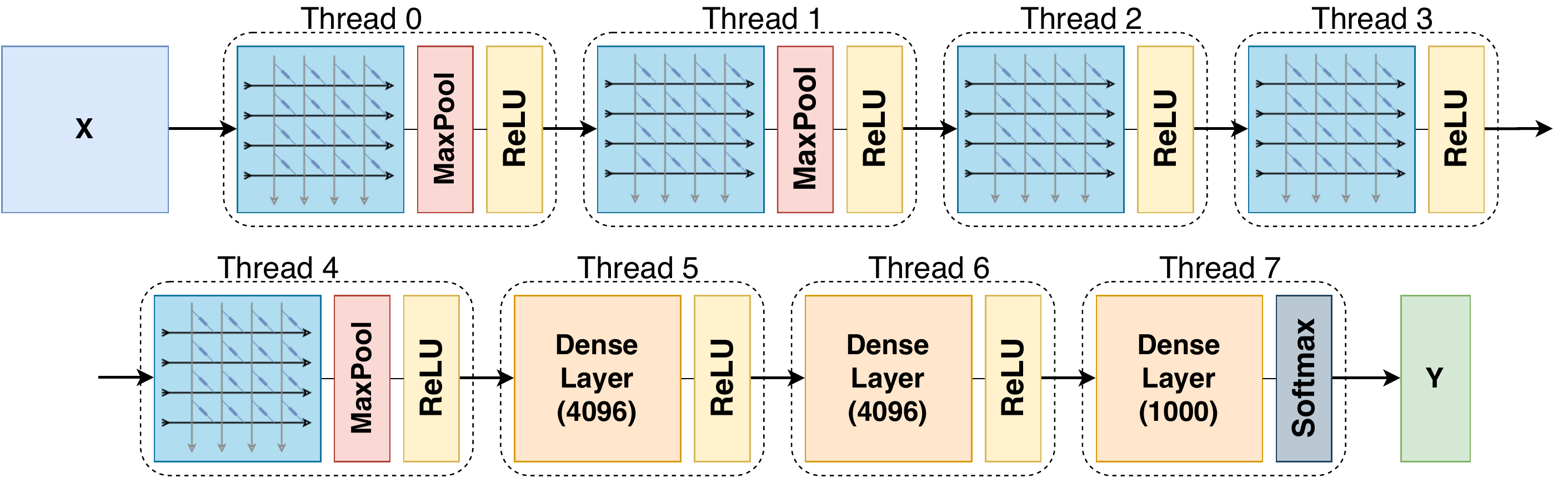}\vspace{0.4cm}
    \textbf{(b) CNN Parameters}\vspace{0.4cm}
    \footnotesize
    \begin{tabular}{|C{1cm}||C{2cm}|C{2cm}|C{2cm}|}
        \hline 
        Layer & CNN-F & CNN-M & CNN-S \\ \hline \hline
        
        Input & \multicolumn{3}{c|}{224x224x3} \\ \hline
        
        conv1 & 64 11x11 kernels & 96 7x7 kernels & 96 7x7 kernels \\
        & stride 4, pad 0 & stride 2, pad 0 & stride 2, pad 0 \\
        & x2 pool, LRN & x2 pool, LRN & x3 pool, LRN \\ \hline
        
        conv2 & 256 5x5 kernels & 256 5x5 kernels & 256 5x5 kernels \\
        & stride 1, pad 1 & stride 1, pad 1 & stride 1, pad 1 \\
        & x2 pool, LRN & x2 pool, LRN & x2 pool \\ \hline
        
        conv3+4 & 256 3x3 kernels & 512 3x3 kernels & 512 3x3 kernels \\
        & stride 1, pad 1 & stride 1, pad 1 & stride 1, pad 1 \\ \hline
        
        conv5 & 256 3x3 kernels & 512 3x3 kernels & 512 3x3 kernels \\
        & stride 1, pad 1 & stride 1, pad 1 & stride 1, pad 1 \\
        & x2 pool & x2 pool & x3 pool \\ \hline
        
        dense1+2 & \multicolumn{3}{c|}{4096 dropout} \\ \hline
        
        dense3 & \multicolumn{3}{c|}{1000 dropout} \\ \hline
        
        \textbf{AIMC params} & 1.7M & 5.6M & 5.5M \\ \hline
        
        % \textbf{Total activations} & 512k & \multicolumn{2}{c|}{1.6M} \\ \hline
    \end{tabular}
    
    \caption{(a) The architecture of the CNNs presented in \cite{chatfield2014return} and their mapping onto the ALPINE systems. The blue boxes with the AIMC tiles represent the convolutional layers. The dense layers are not mapped to AIMC tiles. (b) shows the dimensions and parameters of each CNN. The CNN has 5 convolutional layers (3 with Max Pooling), 3 dense layers, and ReLU activation functions for all layers except the last layer, which uses Softmax.}
    \vskip -1.5em
    
    \label{fig:8-cnn-cases}
\end{figure}

Aggregate results for all of our LSTM experiments are shown in Fig. \ref{fig:7-lstm-aggregate-results}, including multi-core results, results for both system configurations, and factor improvements with the largest of the networks ($n_h = 750$).  When $n_h$ is 256, we observe 1.0-1.5x factor improvements across all metrics and system configurations in the AIMC tile-enabled cases over the digital case due to the very small working set.  However, when $n_h$ increases to 512 and then 750, the run time and energy of the digital application increases up to 9.4x/9.3x with the working set size increase of 7x.  This is in comparison to the AIMC tile-enabled applications, which sees an average run time and energy increase of 1.4x, suggesting a sub-quadratic increase in run-time complexity with higher space complexity.

Relative to the first case study with the MLP, we see smaller maximum time and energy gains (9.4x/9.3x versus 12.8x/12.5x); this is expected as a greater proportion of the LSTM total run-time is dedicated to digital operations that do not see a reduced computational complexity with the introduction of the AIMC tile.  However, these digital operations do see mild performance improvements as a result of lower memory intensity and therefore less cache thrashing (due to space freed from lack of weights loaded in AIMC-enabled LSTMs).

\subsection{Multi-Core Results and Analysis}
Similar to the single-CPU core cases, the multi-CPU core cases also have significant performance gains as $n_h$ grows larger, in comparison to the digital implementation (Fig. \ref{fig:7-lstm-aggregate-results}). By examining the sub-ROIs of the LSTM inference, as seen in Fig. \ref{fig:7-lstm-subroi}, we see that the new bottleneck of the LSTM algorithm is the cell layer dequeuing and activation functions (up to 81.8\% of the inference run time), followed by the cell gate combinations (up to 14.9\% of the inference run time). After more analysis, the activation functions in the cell layer alone account for approximately 70\% of the cell dequeue and activation run time, meaning that the cell layer's digital component in the AIMC tile-enabled systems accounts for up to 57.3\% of the algorithm's run-time.  Hence, unlike with the MLP study, going from single-CPU core to multi-CPU core with the LSTM in the AIMC tile-enabled implementations does result in a speedup of 10\% (cases 1 vs. 4), due to the LSTM cell's parallelized linear operations (but not more due to inter-layer communication).

% Moved these figures here to coax them into being on the correct page.
\begin{figure*}
    \centering
    \includegraphics[width=16.8cm]{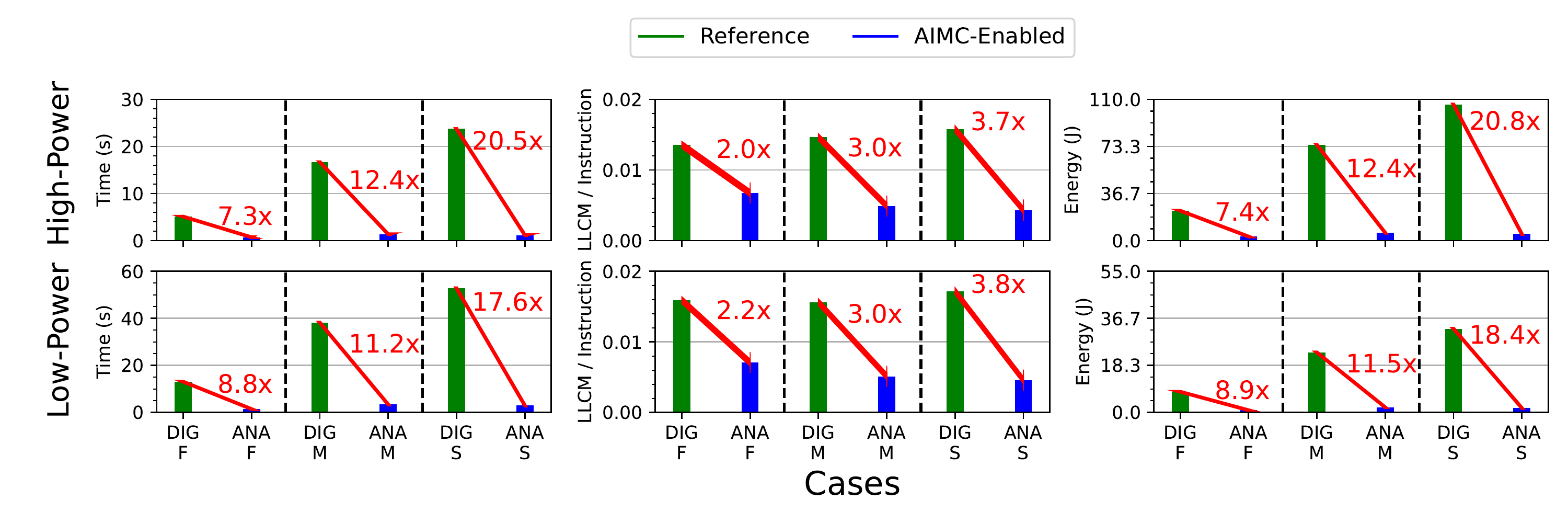}
    \vskip -1.5em
    \caption{Aggregate results for CNN experiments.  From left to right, each column contains total time, memory intensity, and energy results for the High-Power system (top row) and Low-Power system (bottom row) configurations. \textit{ANA} refers to analog AIMC-enabled applications with CNN names corresponding to those in Table \ref{fig:8-cnn-cases}, while \textit{DIG} refers to a digital reference, non-AIMC-enabled, implementation.  The CNNs F, M, and S represent fast, medium, and slow variations on the same CNN architecture, respectively.}
    \label{fig:8-cnn-aggregate-results}
\end{figure*}
\begin{figure}
    \centering
    \vskip -1.5em
    \includegraphics[width=8.2cm]{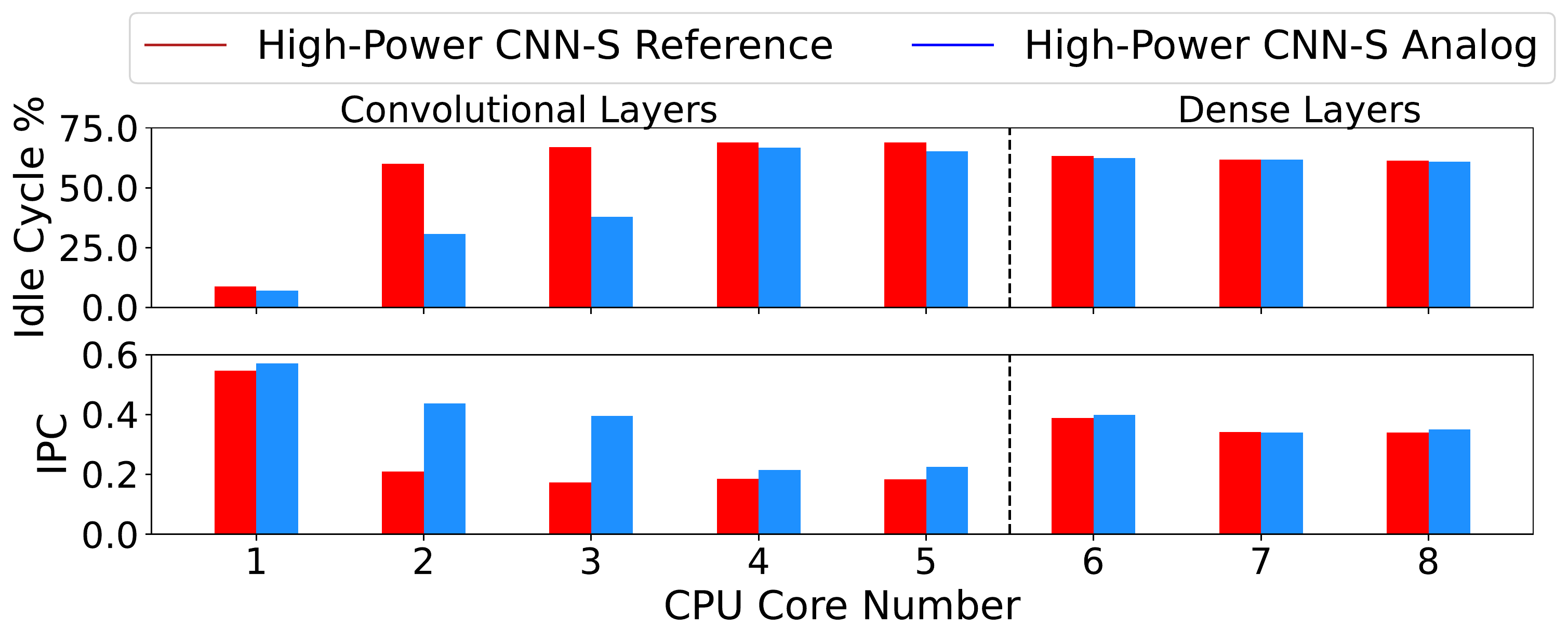}
    \caption{CPU utilization for CNN-S in the high-power system, expressed as the percentage of idle CPU cycles (top) and instructions/cycle (IPC) (bottom).}
    \vskip -1.5em
    \label{fig:8-cnn-utilization-results}
\end{figure}

\subsection{LSTM Complexity}
In the digital reference application, the computation of the LSTM cell outputs involve four MVM operations of quadratic complexity ($4O((n_h) * (x + n_h)) \approx 4O(n^2)$) and nine operations of linear complexity (sigmoid, hyperbolic tangent, array multiplication, array addition; $9O(n_h) \approx 9O(n)$).  In addition to the LSTM cell's added complexity, the softmax operation is used as the fully-connected layer's activation function which doubles in complexity when compared to ReLU ($2O(y) \approx 2O(n)$).  This is in addition to loading inputs and storing results ($O(x) + O(y) \approx 2O(n)$).  Thus the total computational complexity of the ROI of our LSTM application for $N_{inf}$ inferences is $ N_{inf}*(5O(n^2) + 13O(n) \approx O(N_{inf}n^2)$.  Even though the computational complexity of the LSTM is similar to the MLP, it is actually more than 3x more intensive than the MLP in terms of linear digital operations and contains three more MVM operations per inference.

With the introduction of the AIMC tiles in the single- and dual-CPU core cases however, the application queues the concatenated input [$h_{t-1}$, $x$] into the AIMC tile's input memory for the LSTM cell, and then perform all four MVM operations with one CM\_PROCESS instruction call by tiling the LSTM cell weights next to each other (in cases 1 and 2).  Then when the application fetches the result, it is actually fetching the concatenated MVM results of the forget, input, activation, and output gates (MVM results using $W_f$, $W_i$, $W_a$, and $W_o$, respectively) before the activation functions are performed.  Thus the computational complexity of the LSTM cell layer reduces the $4O(n^2)$ factor to $O(1)$ while adding the queuing and dequeuing complexities of $O([h_{t-1}, x]) + O(n_h) \approx 2O(n)$.  Subsequently, the total computational complexity can be reduced to $N_{inf}*(2O(1) + 15O(n)) \approx O(N_{inf}n)$, or linear complexity.  The reduced complexity explains why as $n_h$ increases there is not a substantial increase in the run time of the AIMC-enabled LSTM application, as opposed to non-accelerated (digital reference) mappings.

Similar to the MLP, when the layers are pipelined together in the multicore cases the complexity of both the AIMC-enabled and reference applications reduces by the number of hardware threads used.  Unlike the MLP however, there is not a slowdown as a result of taking our AIMC-tiled application multicore, due to the increased number of digital operations in the LSTM cell being more adequately split across cores.  

\subsection{LSTM Working Set Analysis}
Just like with the MLP application however, the LSTMs also make significant gains in memory intensity by using stationary weights, and thus reducing the size of the working set.  If we calculate the approximate working set of the LSTM application, it is comprised of the input ($x + h_{0}$), the LSTM cell layer's weights ($4 * (n_h * (n_h + x))$, the fully-connected layer's weights ($n_h * y$), the intermediary result in-between the layers ($n_h$), and the output ($y$).  The total size complexity of the LSTM application per inference using 8-bit types is $(x + n_h) + 4(n_h^2 + n_hx) + n_h + n_hy + y$.  Using the numbers from Table \ref{tab:7-lstm-params}, the size of the working set for $n_h$ = 256, 512, 750 is 378kB, 1.28MB, and 2.59MB, respectively.  Even in the smallest variant of the LSTM application, the working set cannot entirely fit in the private caches of the CPU core(s).  When $n_h$ is 512 or 750, in both low-power and high-power system configurations, the CPU core(s) must go out to L2 cache and main memory to contain the working set.

When we utilize AIMC tiles in our LSTM application, the weights are removed from the working set because they are never needed by the CPU core(s) during the ROI, thus reducing our total size complexity to $(x + n_h) + n_h + y$, which is 0.66kB, 1.17kB, and 1.65kB for $n_h$ = 256, 512, and 750, respectively.  For all values of $n_h$ tested, the working set can fit entirely in L1 private caches for both low-power and high-power system configurations, hence the increasing performance improvements and lower memory intensity with greater values of $n_h$.

Therefore, we conclude that while AIMC tiles greatly speedup neural networks where MVMs are the dominating operation, careful attention must be paid to both the size of the neural network and the proportion of other non-optimized digital operations.  This case study of the LSTM shows that when linear operations are more dominant, that realized gains in performance are lower with the AIMC tiles (as compared with the first exploration study), and thus these other linear operations which are not optimized with the inclusion of the AIMC tiles become the new bottleneck.  Furthermore, when the neural network is small enough to efficiently leverage a CPU core's L1 private cache and a small portion of L2 cache (as when $n_h$ is 256), the overhead of queuing inputs and dequeuing outputs to and from the AIMC tile ends up having a similar run-time (with only minor performance gains) to the reference application, which otherwise invalidates or diminishes the potential benefit of introducing AIMC tiles.

\section{Exploration Three: Convolution Neural Networks}
\subsection{CNN Architecture}
For our last exploration study, we explore the benefits of introducing AIMC tiles for CNNs in an 8-CPU core MPSoC. Contrary to the previous studies, where each layer weights are used only once per inference, convolution operations require multiple passes on weights per inference via shifting kernels over the feature maps. To perform the convolution operations in AIMC tile-enabled applications, we flatten the kernels into columns and store these in the columns of the AIMC tile, as described in \cite{yakopcic2016memristor, Y2020joshiNatComm}. The feature maps are also flattened and queued to the AIMC tile. 

We explore the three CNN variants presented in \cite{chatfield2014return} to act as a baseline for modern CNNs; they are labeled \textit{CNN-F(ast)}, \textit{CNN-M(edium)}, and \textit{CNN-S(low)}.  While CNN-S and CNN-M are similarly sized, the increase of the MaxPool operation factor in layers 1 and 5 from x2 to x3 increases the computational requirements of CNN-S significantly by performing strided 3x3 pool operations instead of 2x2 pool operations. Figure \ref{fig:8-cnn-cases} (a) shows the proposed CNN implementation and data flow, while Figure \ref{fig:8-cnn-cases} (b) reports the CNN architecture parameters. Fine-grained pipelining is applied for the data-flow; the convolutions are performed whenever the corresponding input volume of the feature map is available. Contrary to the previous exploration studies, we utilize the AIMC tiles only for convolutional layers. The feed-forward layers are processed in the CPU; these layers are executed only once as opposed to the convolutional layers and therefore do not constitute a bottleneck. 

\subsection{Results and Analysis}
We present our results in Fig. \ref{fig:8-cnn-aggregate-results}.  The largest performance increase with respect to the CPU-only implementation is recorded for the largest CNN variant "S". This configuration exhibits the maximum speedup of 20.5x, a memory intensity improvement of 3.7x, and an energy improvement of 20.8x for the high-power system.

While in the prior case studies the core utilization across layers does not vary widely across experiments, the CNN benchmark is used to examine the AIMC acceleration with uneven CPU core utilization. To this end, Figure \ref{fig:8-cnn-utilization-results} shows both the CPU idle percentage and the instructions per cycle (IPC) count for each individual CPU core in our high-power CNN-S application. CNN-M and CNN-F in both low-power and high-power system configurations exhibit very similar trends. The utilization of the first convolutional layer is similar in both CPU-only and AIMC tile-enabled benchmarks due to input load from memory. For convolutional layers 2 and 3, AIMC tiles provides significant benefits with idle cycles decreasing up to 4x. Likewise, IPC increases relative to the CPU-only benchmark by up to 3x. Convolutional layers 4 and 5 exhibit more idle cycles in the AIMC-based implementation in comparison to prior layers. This is partly attributed to reduced size of the feature maps, owing to the stride and pooling operations of the previous layers. The fully-connected layers' CPU cores spent the most time idling.% as fully-connected layers are utilized once during inference.
%Yet, another factor impacting the 

%As it is already established that the memory intensity of the AIMC-enabled application goes down significantly, the drop in IPC for these last two convolution layers is direct evidence of a communications and computational bottleneck.  Further evidence comes from how the fully-connected layers (which are not accelerated) do not incur a speedup and subsequently do not see an increase in IPC, despite more system resources (namely, L2 shared cache) being made available by the lack of CNN kernels in the working set in the AIMC tile-enabled CNN.

Owing to the fine-grained activation pipelining, the amount of data to be communicated between layers are significantly reduced. Yet, the total inference run-time is more than the MLP and LSTM cases owing to the multiple passes over the convolutional kernels.

As a result, while the AIMC tiles offer significant speedups for CNNs as well, further exploration is needed to optimize the data flow for shifting bottlenecks in layers. This includes replicating the initial layer convolutional kernels to balance the CNN pipeline owing to varying feature map sizes \cite{Y2021dazziFCN}, including local SRAM in the AIMC tile for avoiding queueing the same input volume of the feature maps multiple times \cite{Y2021dazziFCN} and minimizing core-to-core communication overheads. %We would like to note that all these explorations 
This investigation can be carried out on the ALPINE framework.

\section{Conclusion}
In this work, we presented and explored the performance benefits of a novel architecture that utilizes tightly-coupled AIMC tiles.  We implemented ALPINE, a gem5-X extension for modeling AIMC tiles in the gem5 full system simulation framework.  We extended the ARMv8 ISA with custom instructions that interface AIMC tiles directly from their execution in the CPU.  For ease of programming and using AIMC tiles, we implemented a dedicated software library (AIMCLib).  Using ALPINE and AIMClib, we then implemented and tested three different exploration studies across two system configurations, namely, single and multi-core MLPs, single and multi-core LSTMs, and finally multi-core CNNs.  Through these exploration studies we observed how computational and size complexity is reduced by leveraging the AIMC tiles, ultimately demonstrating up to  20.5x/20.8x performance/energy gains with respect to a SIMD-enabled fully-digital reference implementation.

\section*{Acknowledgments}
We thank Geethan Karunaratne, Pier Andrea Francese and Riduan Khaddam-Aljameh for technical discussions. This  work  has  been  supported  by  the  EC  H2020 WiPLASH  (GA  No.  863337) project and the ERC Consolidator Grant COMPUSAPIEN (GA No. 725657) projects.  We also thank Marceline and Lohse Klein for their support.

\bibliographystyle{IEEEtran}
\bibliography{library}

\section*{Author Biographies}
\newcommand{\biowidth}{0.9in}
\newcommand{\bioheight}{1.11in}

\vskip -3.0em
\begin{IEEEbiography}[{\includegraphics[width=\biowidth,height=\bioheight,clip,keepaspectratio]{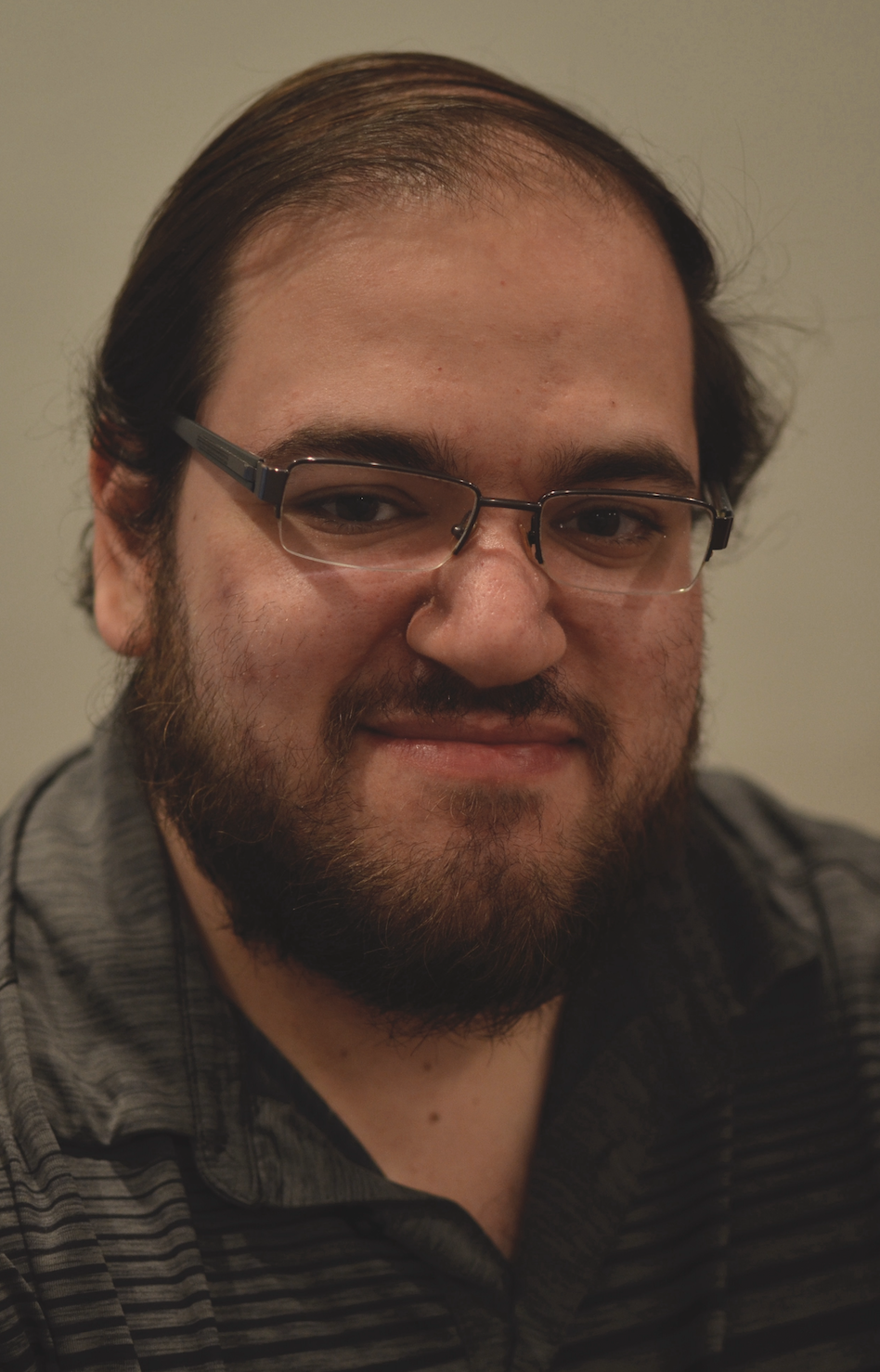}}]{Joshua Klein}
is a 3rd-year doctoral student at the Embedded Systems Laboratory of the École Polytechnique Fédérale de Lausanne (EPFL), Switzerland. He received his B.Sc. in Computer Engineering in 2017 magna cum laude and his M.Sc. in Electrical and Computer Engineering in 2019 from Boston University, USA.  His research interests include system-level modeling and simulation, RISC architectures, and novel accelerators for machine learning applications.
\end{IEEEbiography}
\vskip -3.0em

\begin{IEEEbiography}[{\includegraphics[width=\biowidth,height=\bioheight,clip,keepaspectratio]{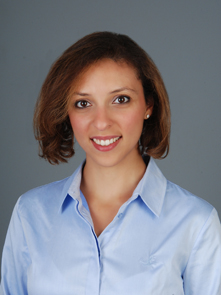}}]{Irem Boybat}
is a Research Staff Member at IBM Research – Zurich. She received her B.Sc. degree in Electronics Engineering from Sabanci University, Turkey (2013), and M.Sc. and Ph.D. degrees in Electrical Engineering from École Polytechnique Fédérale de Lausanne (EPFL), Switzerland (2015 and 2020, respectively). Her research interests include in-memory computing for AI systems, neuromorphic computing, and emerging resistive memory. %She has co-authored over 40 scientific papers in journals and conferences, received three best conference presentation/paper/poster awards and holds 5 granted patents. She was a co-recipient of the 2018 IBM Pat Goldberg Memorial Best Paper Award, 2018 IBM Research Division Award on neuromorphic computing using phase-change memory devices, and 2020 EPFL PhD Thesis Distinction in Electrical Engineering.
\end{IEEEbiography}
\vskip -3.0em

\begin{IEEEbiography}[{\includegraphics[width=\biowidth,height=0.9in,clip,keepaspectratio]{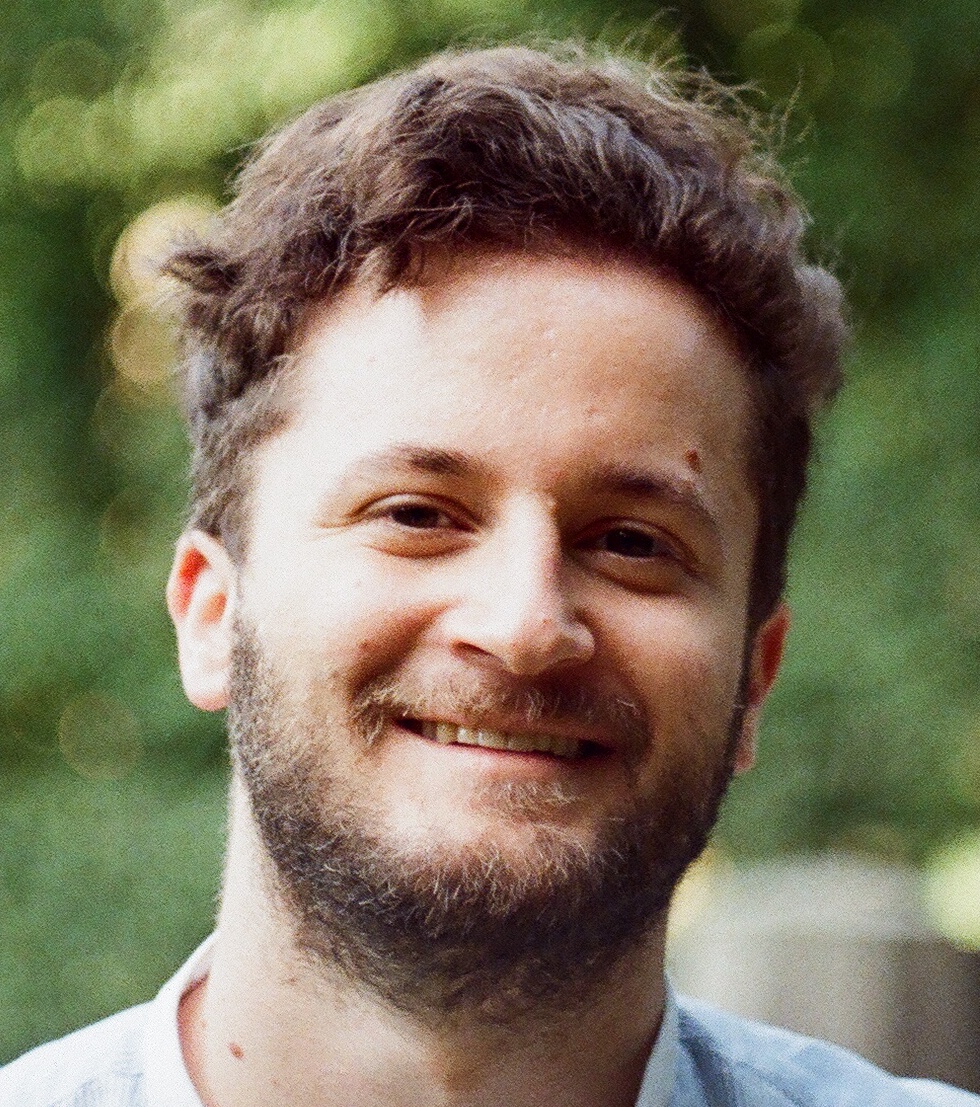}}]{Martino Dazzi} received his B.Sc. and M.Sc. (cum laude) degrees in Electronic Engineering from the University of Udine, Italy, in 2015 and 2017, respectively. 
%Between 2017 and 2018, he held a Research Assistant position at Alma Mater Studiorum-Universita’ di Bologna, Italy, focusing on mixed-signal cicuit design. Between 2018 and 2021, he held a research position at IBM Research Zurich, Switzerland. In 2021 he received his Ph.D. degree in Electrical Engineering from ETH Zurich, Switzerland.
In 2021, he co-founded Axelera AI, where he currently holds a position as Algorithm and Quantization Researcher. His main research interest is in machine learning, specifically in energy-efficient, reduced-precision design of neural networks.
\end{IEEEbiography}

\begin{IEEEbiography}[{\includegraphics[width=\biowidth,height=\bioheight,clip,keepaspectratio]{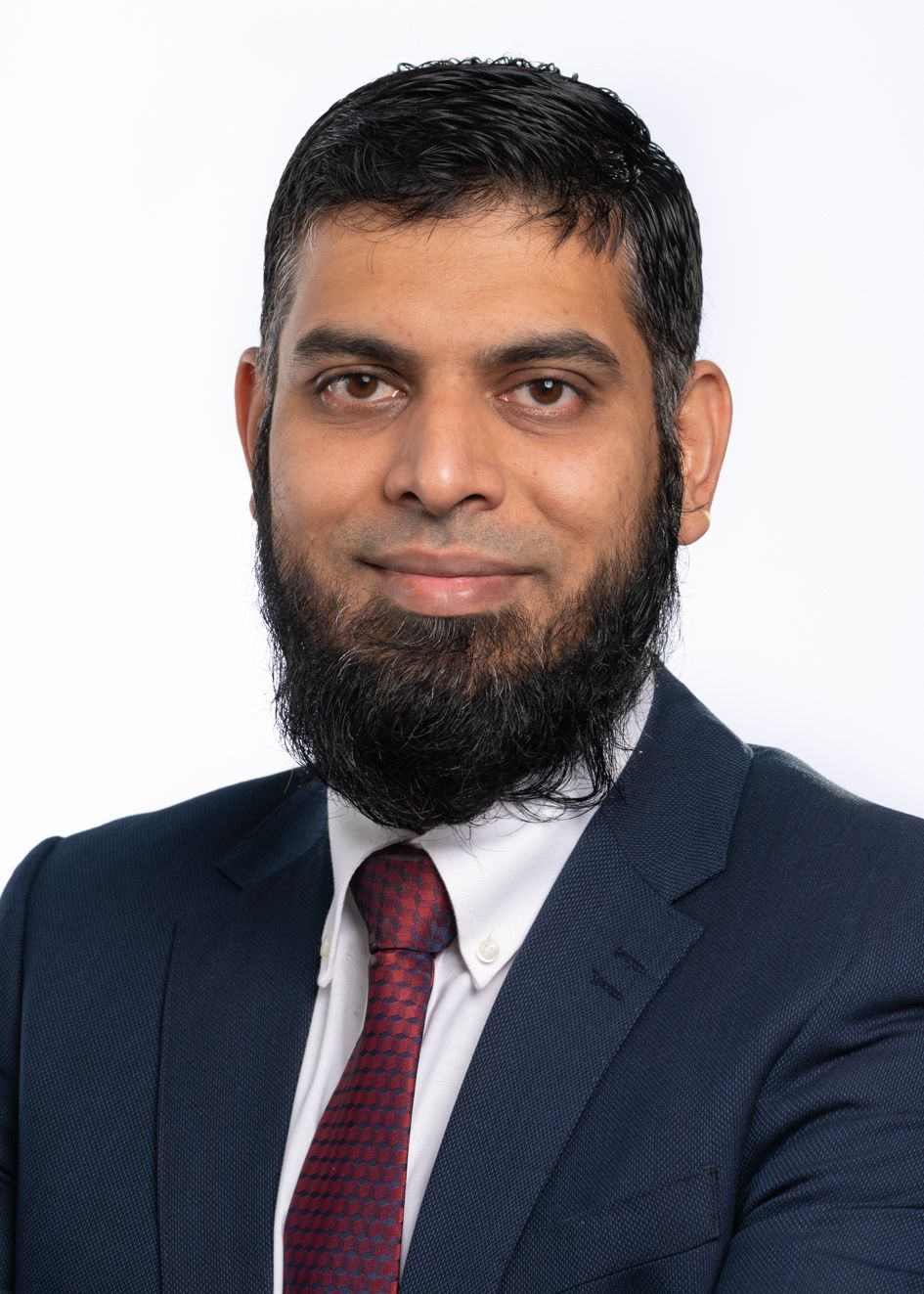}}]{Yasir Mahmood Qureshi} received his Ph.D. in Electrical Engineering from EPFL, Switzerland, in 2021. He is currently working as a Staff Design Engineer at Infineon Technologies, Ireland. His research interests include energy-efficient computing, heterogeneous compute and hybrid memory architectures, compute sub-systems for automotive microcontrollers, and safety-critical and secure compute architectures.\end{IEEEbiography}
\vskip -3.0em

\begin{IEEEbiography}[{\includegraphics[width=\biowidth,height=\bioheight,clip,keepaspectratio]{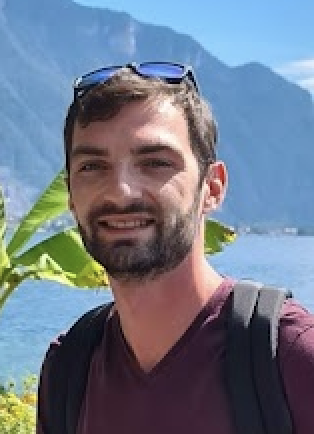}}]{Alexandre Levisse}
received his Ph.D. degree in Electrical Engineering from CEA-LETI, France, and from Aix-Marseille University, France, in 2017. From 2018 to 2021, he was a post-doctoral researcher in the Embedded Systems Laboratory at the Swiss Federal Institute of Technology Lausanne (EPFL). From 2021, he works as a scientist in EPFL. His research interests include circuits and architectures for emerging memory and transistor technologies as well as in-memory computing and accelerators.\end{IEEEbiography}
\vskip -3.0em

\begin{IEEEbiography}[{\includegraphics[width=\biowidth,height=\bioheight,clip,keepaspectratio]{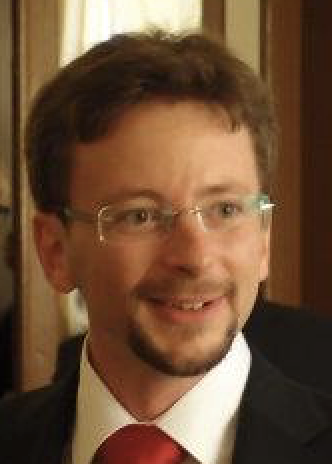}}]{Giovanni Ansaloni}
is a researcher at the Embedded Systems Laboratory of EPFL (Lausanne, CH). He previously worked as a Post-Doc at the University of Lugano (USI, CH) between 2015 and 2020, and at EPFL between 2011 and 2015. He received a Ph.D. degree in Informatics from USI in 2011. His research efforts focus on domain-specific and ultra-low-power architectures and algorithms for edge computing systems, including hardware and software optimization techniques.\end{IEEEbiography}
\vskip -3.0em

\begin{IEEEbiography}[{\includegraphics[width=\biowidth,height=\bioheight,clip,keepaspectratio]{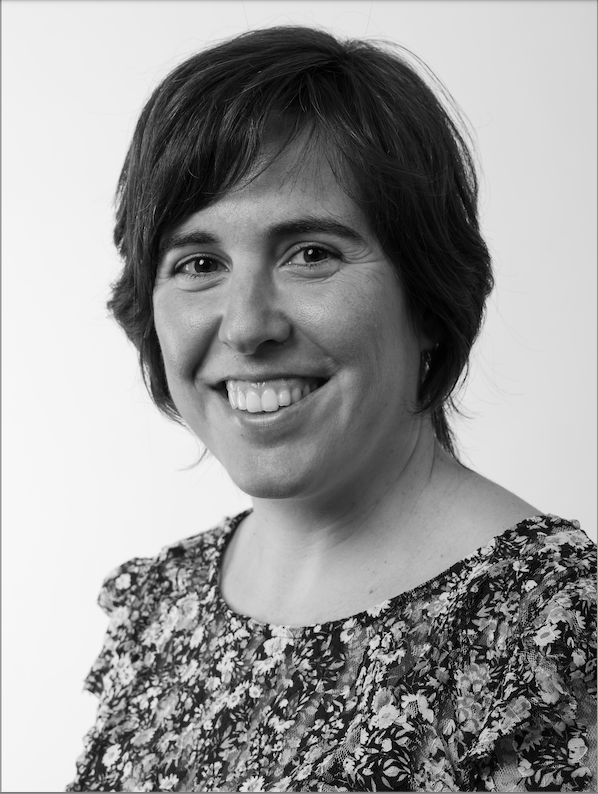}}]{Marina Zapater} is an Associate Professor in the REDS Institute at the School of Engineering and Management of Vaud (HEIG-VD) of the University of Applied Sciences Western Switzerland (HES-SO) since 2020. %She was post-doctoral Research Associate in the Embedded System Laboratory (ESL) at the Swiss Federal Institute of Technology Lausanne (EPFL), Switzerland, from 2016 to 2020. 
She received her Ph.D. degree in Electronic Engineering from Universidad Politécnica de Madrid, Spain, in 2015. Her research interests include thermal, power, and performance design and optimization of complex heterogeneous architectures.%, from embedded AI-enabled edge devices to high-performance computing processors; and energy efficiency in servers and data centers. %In these fields, she has co-authored more than 50 papers in top-notch conferences and journals. She is an IEEE and CEDA member.
\end{IEEEbiography}
\vskip -3.0em

\begin{IEEEbiography}[{\includegraphics[width=\biowidth,height=\bioheight,clip,keepaspectratio]{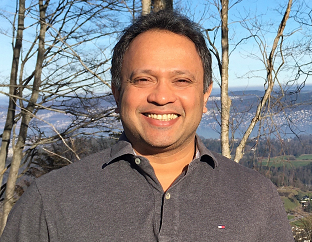}}]{Abu Sebastian}
is a Distinguished Research Staff Member at IBM Research – Zurich. He received a B. E. (Hons.) degree in Electrical and Electronics Engineering from BITS Pilani, India, in 1998 and M.S. and Ph.D. degrees in Electrical Engineering (minor in Mathematics) from Iowa State University in 1999 and 2004, respectively. He manages the research effort on in-memory computing at IBM Research Zurich. %He is the author/co-author of over 200 publications in peer-reviewed journals/conference proceedings and holds over 70 US patents.
%In 2015 he was awarded the European Research Council (ERC) consolidator grant and in 2020, he was awarded an ERC Proof-of-concept grant. He is an IBM Master Inventor since 2016. In 2019 he received the Ovshinsky Lectureship Award for his contributions to "Phase-change materials for cognitive computing". He has served on the technical program committees of several conferences including IEDM, AICAS and E\textbackslash PCOS.
\end{IEEEbiography}
\vskip -3.0em

\begin{IEEEbiography}[{\includegraphics[width=\biowidth,height=\bioheight,clip,keepaspectratio]{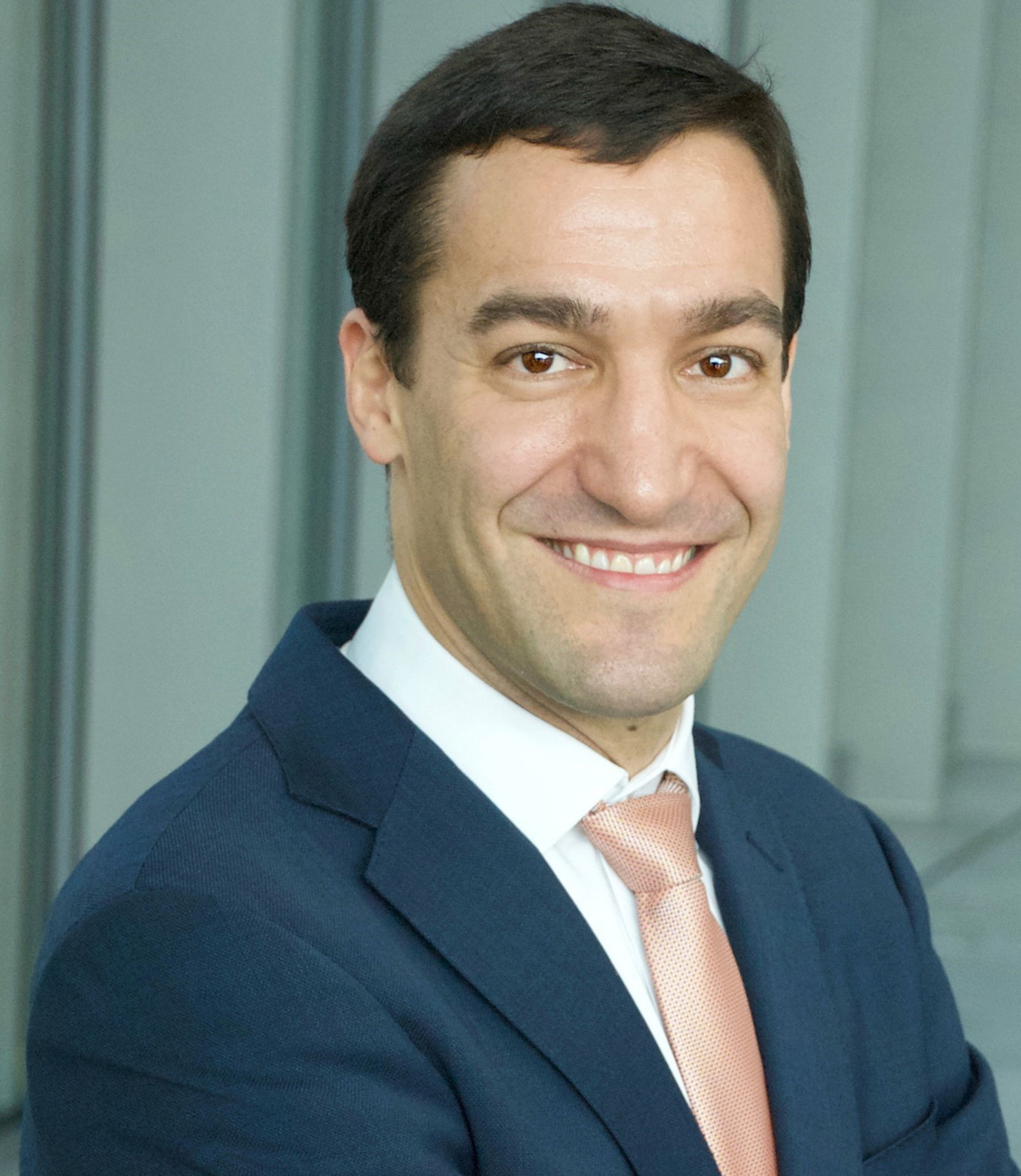}}]{David Atienza}
is a full professor of electrical and computer engineering, and head of the Embedded Systems Laboratory (ESL) at EPFL, Switzerland. He received his Ph.D. in computer science and engineering from UCM, Spain, and IMEC, Belgium, in 2005. His research interests include system-level design methodologies for multi-processor system-on-chip (MPSoC) servers and edge AI architectures. %He has co-authored more than 350 papers, one book, and 12 patents. %Dr. Atienza has received, among other recognitions, the ICCAD 10-Year Retrospective Most Influential Paper Award in 2020, the Most Influential DAC Under-40 Innovators Award in 2018, and an ERC Consolidator Grant in 2016. He is an IEEE Fellow and an ACM Distinguished Member.
\end{IEEEbiography}

\end{document}